\RequirePackage{ifpdf}
\documentclass[hyper,12pt,a4paper]{JHEP3}
\pdfoutput=1
\usepackage{cite}
\usepackage{psfrag}
\usepackage{graphicx}
\usepackage{enumitem}
\usepackage{bm}
\usepackage{amsmath}
\usepackage{float}

\usepackage{amssymb}
\usepackage{fancyhdr}
\usepackage{url}
\usepackage{epstopdf}
\usepackage[utf8]{inputenc}
\usepackage{verbatim}
\usepackage{float}
\usepackage[font=small,labelfont=bf]{caption}
\usepackage{mathrsfs}
\usepackage[parfill]{parskip}
\usepackage{subcaption}
\graphicspath{{chobi/}}

\numberwithin{equation}{section}
\usepackage{tabularx}

\newcommand{\Tr}{\text{Tr}}

\newcommand{\seff}[1]{S_{\text{eff}}[{#1}]}

\newcommand{\ben}{\begin{eqnarray}\displaystyle}
\newcommand{\een}{\end{eqnarray}}

\newcommand{\be}{\begin{equation}}
\newcommand{\ee}{\end{equation}}

\newcommand{\bc}{\begin{center}}
\newcommand{\ec}{\end{center}}

\newcommand{\eesp}{\end{split}}
\newcommand{\bsp}{\begin{split}}

\newcommand{\Rmnum}[1]{\expandafter\@slowromancap\romannumeral #1@}
\renewcommand{\b}{\beta}		%%% Redefinition
\renewcommand{\l}{\lambda}	%%% Redefinition
\newcommand{\m}{\mu}								%%% Useless
\newcommand{\q}{\theta}	
	
\renewcommand{\r}{\rho}		%%% Redefinition
\newcommand{\s}{\sigma}

\newcommand{\bo}{\beta_{1}}

\newcommand{\D}{\Delta}

\renewcommand{\S}{\Sigma}	%%% Redefinition

\newcommand{\cA}{\mathcal{A}}

\newcommand{\cC}{\mathcal{C}}
\newcommand{\cD}{\mathcal{D}}

\newcommand{\cF}{\mathcal{F}}

\newcommand{\cH}{\mathcal{H}}

\newcommand{\cK}{\mathcal{K}}

\newcommand{\cM}{\mathcal{M}}

\newcommand{\cR}{\mathcal{R}}
\newcommand{\cS}{\mathcal{S}}
\newcommand{\cT}{\mathcal{T}}

\newcommand{\cW}{\mathcal{W}}

\newcommand{\cZ}{\mathcal{Z}}

\newcommand{\expa}[1]{\exp\left( #1 \right)}

\newcommand{\zcs}{\cZ_{\text{CS}}}

\newcommand{\sina}[1]{\sin\lb #1 \rb}
\newcommand{\cosa}[1]{\cos\lb #1 \rb}

\newcommand{\ra}{\rightarrow}

\newcommand{\lB}{\left [}
\newcommand{\rB}{\right ]}
\newcommand{\lb}{\left (}
\newcommand{\rb}{\right )}

\newcommand{\where}{\text{where}}
\newcommand{\with}{\text{with}}
\newcommand{\tand}{\text{and}}

\newcommand{\bensp}{\begin{eqnarray}\begin{split}}
\newcommand{\eensp}{\end{eqnarray}\end{split}}

\newcommand{\bnm}{\begin{matrix}}
\newcommand{\enm}{\end{matrix}}
\def\Xint#1{\mathchoice
{\XXint\displaystyle\textstyle{#1}}%
{\XXint\textstyle\scriptstyle{#1}}%
{\XXint\scriptstyle\scriptscriptstyle{#1}}%
{\XXint\scriptscriptstyle\scriptscriptstyle{#1}}%
\!\int}
\def\XXint#1#2#3{{\setbox0=\hbox{$#1{#2#3}{\int}$ }
\vcenter{\hbox{$#2#3$ }}\kern-.6\wd0}}

\newcommand{\half}[1]{{#1\over 2}}

\newcommand{\gt}{\theta}
\newcommand{\gl}{\lambda}

\newcommand{\gb}{\beta}

\newcommand{\stso}{$S^2\times S^1$}

\newcommand\dnorm[1]{\langle#1\rangle}

\newcommand{\cs}{Chern-Simons\ }
\newcommand{\sm}{\cM_{(g,p)} }

\newcommand{\stwoso}{S^2\times S^1 }
\newcommand{\un}{U(N) }

\newcommand{\zcst}{\cZ_{\text{CS}}[S^3,U(N),k] }

\title{Chern-Simons Theory on Seifert Manifold and Matrix Model} 
\author{Arghya Chattopadhyay\footnote{arghya@iiserb.ac.in} $^{1}$,
	 Suvankar 	Dutta\footnote{suvankar@iiserb.ac.in} $^{1}$, Neetu\footnote{neetuj@iiserb.ac.in} $^{1}$\\
	$^{1}$Department of Physics \\
	Indian Institute of Science Education and Research Bhopal \\
	Bhopal 462 066, 
	India\\
}
\abstract{
	Chern-Simons (CS) theories with rank $N$ and level $k$ on Seifert manifold are discussed. The partition functions of such theories can be written as a function of modular transformation matrices summed over different integrable representations of affine Lie algebra $u(N)_k$ associated with boundary Wess-Zumino-Witten (WZW) model. Using properties of modular transform matrices we express the partition functions of these theories as a unitary matrix model. We show that, the eigenvalues of unitary matrices are discrete and proportional to hook lengths of the corresponding integrable Young diagram. As a result, in the large $N$ limit, the eigenvalue density develops an upper cap. We consider CS theory on $S^2\times S^1$ coupled with fundamental matters and express the partition functions in terms of modular transformation matrices. Solving this model at large $N$ we find the dominant integrable representations and show how large $N$ representations are related to each other by transposition of Young diagrams as a result of level rank duality.
	
	Next we consider $U(N)$ CS theory on $S^3$ and observed that in Seifert framing the dominant representation is no longer an integrable representation after a critical value of 't Hooft coupling. We also show that CS on $S^3$ admits multiple (two-gap phase) large $N$ phases with the same free energy.
}

\begin{document}
%\fi

\section{Introduction}\label{sec:introduction}

Study of topological objects in physics is an extremely interesting subject. One of the earliest examples of topological objects is the Dirac monopole \cite{diracmonopole}. Today we know a large number of such examples in physics starting from quantum mechanics to string theory. An example of such objects in mathematics is \emph{knot} (closed path). Mathematically, a knot is just a smooth closed, non-self-intersecting curve in three dimensions. Knot theory turns out to be specially useful to study the physics of two-dimensional many body systems\cite{wj8,wj11}. A field theoretic realization of knots and links (collection of non-intersecting knots) was discovered by Witten in his groundbreaking work in 1989 \cite{wittenjones}. In that paper he showed that physical observables (Wilson loops) in Chern-Simons gauge theory in three dimensions are related to knot polynomials in the same dimensions and thus opened up a plethora of new possibilities for both mathematicians and physicists.

To discuss the connection in a little detail, we start with pure Chern-Simons theory with gauge group $G$ and level $k$ on a compact three manifold $\mathbb{M}$. The action is given by 
\begin{equation}\label{eq:csaction}
S_{CS}[\mathbb{M},G,k]={k\over 4\pi}\int_\mathbb{M} \Tr\left( A\wedge dA+{2\over 3}A\wedge A\wedge A\right).
\end{equation}
The action does not require any metric on $\mathbb{M}$. This defines the most simple version of \emph{topological field theories}, namely, the \emph{Schwarz type topological field theory}. The partition function for Chern-Simons theory on $\mathbb{M}$
\begin{equation}
\zcs[\mathbb{M},G,k]=\int [DA] e^{iS_{CS}}
\end{equation}
itself defines a topological invariant of the manifold $\mathbb{M}$. To explain the relation between observables in Chern-Simons theory and knot polynomials we consider Wilson loop operators in representation $R$ of $G$ along an oriented knot $\mathcal{K}$ in $\mathbb{M}$, defined as
\begin{equation}
\mathcal{W}_R^\mathcal{K}(A)=\Tr_R U_\mathcal{K} \quad \text{with } U_\mathcal{K}=P \exp \oint_\mathcal{K} A
\end{equation}
where $U_\mathcal{K}$ is called holonomy around the knot $\mathcal{K}$. Since Wilson loop operators are gauge invariant by definition and in this particular case they are metric independent as well, it is quite obvious that the correlation functions of the Wilson loop operators 
\ben
\langle \cW_{R_1,\cdots R_n}^{\cK_1,\cdots, \cK_n}\rangle = \int [DA] e^{iS_{CS}} \prod_{i=1}^{n} \cW_{R_i}^{\cK_i}(A)
\een
generate topological invariants of the theory. Witten\cite{wittenjones} proved that these topologically invariant correlation functions are precisely the knot invariants\footnote{For $G\equiv U(N)$ or $G\equiv SU(2)$ or $G\equiv SO(N)$ this correlation functions are in turn related to HOMFLY polynomial or the Jones polynomial or Kauffman polynomials respectively.}.

There is a subtle caveat in this seemingly simple story. The metric independence of the classical Lagrangian does not trivially generalise to the quantum version of the theory. Witten showed that the quantum version of Chern-Simons theory preserves topological invariance but at the expense of a choice of \emph{``framing"}. Correlation functions of Wilson loop operators along different knots, depend on \emph{linking number} between the knots involved in the computation \cite{wittenjones,Marino:2004eq}. The linking number between two knots is a topologically invariant quantity; therefore, in general the correlation functions are also topologically invariant. But the subtlety arises when different Wilson loop operators are taken along the same knot. Then the notion of framing becomes important because in general the self-linking number\footnote{Also known as \emph{cotorsion} or the \emph{writhe}.} of a knot $\mathcal{K}$ is not a topologically invariant quantity. Therefore to preserve topological invariance of the correlation function one needs to modify the definition of self-linking number by a choice of framing\cite{wittenjones,Marino:2004eq}. To generate a consistent notion of self-linking number one defines another knot $\mathcal{K}_f$ around $\mathcal{K}$ specified by some normal vector field $n$ and defines self-linking number as the linking number between $\mathcal{K}$  and $\mathcal{K}_f$. This manner of regularisation reinstates topological invariance of correlation functions but at the expense of its dependence on some integer $p$ defining the linking number between $\mathcal{K}$  and $\mathcal{K}_f$. To visualise the situation, we can imagine the set of normal vectors defined by the vector field $n$, as a tangled ribbon defining an orientable surface, the sides of which are bounded by $\mathcal{K}$ and $\mathcal{K}_f$. In principle there are many ways to construct such ribbons so that one side always coincides with $\mathcal{K}$, and each of this choices renders different framings of the knot. The \emph{canonical framing} is defined as some choice of $n$ such that the self-linking number  $p$ is zero. The framing which is very crucial for this paper is called the \emph{Seifert framing}, where the knot is pushed along the Seifert surface\footnote{A compact, connected, oriented surface embedded in the three manifold having the knot $\mathcal{K}$ as its boundary such that the orientation of the knot is consistent with its own.} to generate $\mathcal{K}_f$ for regularisation. Now it is obvious that a change of framing is nothing but changing the choice of the vector field $n$, which just renders a change in the value of the integer $p$ defined above. Following \cite{wittenjones} it can be shown that, under a change of framing of $\mathcal{K}_i$ by $p_i$ units, the correlation function of the Wilson loops changes as 
\ben
\langle \cW_{R_1,\cdots R_n}^{\cK_1,\cdots, \cK_n}\rangle \ra  \exp\lb 2\pi i \sum_{i=1}^{n} p_i h_{R_i}\rb \langle \cW_{R_1,\cdots R_n}^{\cK_1,\cdots, \cK_n}\rangle
\een
where, $h_{R_i} = \frac{C_{R_i}}{2(k+N)}$ with $c_{R_i}$ being the quadratic Casimir in the representation $R_i$ of $G(N)$.

Not only the correlators, but also the partition function depends on choice of framing. Atiyah\cite{atiyah1990} showed that for every three manifold $\mathbb{M}$ different framing choices can be labeled by an integer $s\in\mathbb{Z}$ ($s$ is self-linking number) such that the canonical framing is given by $s=0$. As a result, if two framings differ by an integer $s$,  the corresponding partition functions are related by \cite{Marino:2004eq}
\begin{equation}\label{eq:framedepend}
\zcs[\mathbb{M},G,k]=\exp\left({\pi i s c\over 12}\right)\zcs[\mathbb{M},G,k]; \quad c={kd\over k+y}
\end{equation}
where $d$ and $y$ are the dimension and dual Coxeter number of the group $G$, and $c$ is the central charge of the Wess-Zumino-Witten (WZW) model\footnote{See \cite{yellowbook} for a pedagogical review of WZW and \cite{wittenjones} for a detailed relation between \cs theory and WZW model.} with the affine gauge group $G_k$. One should realise how WZW model naturally arises in the context of quantizing Chern-Simons theory on three manifolds. The CS/WZW correspondence is in some sense a predecessor of AdS/CFT. The status of AdS/CFT is still at the level of a conjecture whereas one can show that Chern-Simons theory quantised on a closed three manifold can be described exactly by a two dimensional WZW model\footnote{For a comparison between the CS/WZW with $AdS_3/CFT_2$ one can look at the beautiful paper by Gukov et al \cite{Gukov:2004id}.}. Since the CS/WZW is an exact correspondence, one can write observables in \cs theory in terms of objects in the \emph{dual} WZW model\footnote{The central statement which connects these two completely different theories in two different dimensions is that the Hilbert space that one gets while quantising the $(2+1)$ dimensional CS theory turns out to be the space of conformal blocks for the $(1+1)$ dimensional WZW theory. For example if one starts quantising a pure level $k$ CS theory with gauge group $G$ on $\Sigma_g\times S^1$, then the \emph{physical} Hilbert space of CS theory $H_\Sigma$, turns out to be finite dimensional and can be described as the space of conformal blocks of a WZW theory on $\Sigma_g$ with the affine gauge group $G_k$. As spelled out by\cite{wittenjones} one can generalise this result to a generic Seifert manifold $\mathcal{M}_{(g,p)}$ (a circle bundle over $\Sigma_g$ with first Chern class $p$. Ex. $S^3/\mathbb{Z}_p$ for $(g,p)=(0,p)$)  by doing surgeries over $\Sigma_g\times S^1$. In fact different choices of doing surgery to go from one manifold to the other, results into generating the same partition function but in different choice of framings. Lot of work has been done on \cs theories on Seifert manifolds, most recent of which is \cite{Blau:2018cad} and one can look at the references therein for other related works.}. These objects are modular transformation matrices\cite{yellowbook} of the affine lie algebra. This precise relation is our starting point in this paper. We illustrate this relation in detail in the subsequent section.

In this paper we study large $N$ properties of \cs theory with level $k$ and gauge group $G=SU(N)$ or $U(N)$\footnote{In this paper we work in large $N$ limit. Our results are not sensitive to this choice.} on Seifert manifold $\mathcal{M}_{(g,p)}$. The partition function for the same in Seifert framing can be written as a function of modular transform matrices\footnote{The characters of the integrable representations of an affine Lie algebra $su(N)_k$ transform into one another under modular transformations. The two generators of this modular group are conventionally denoted by $\mathcal{S}$ and $\mathcal{T}$. See section \ref{sec:csonSM} for details.} of affine Lie group summed over \emph{integrable} representations. An integrable representation of $u(N)_k$ (or $su(N)_k$) has maximum $k$ columns and $N$ rows in Young diagram. The restriction on representations follows from the fact that there is a precise relation between Hilbert space $\cH_{\Sigma}$ of \cs theory with level $k$ and gauge group $SU(N)$ on $\Sigma\times S^1$ and WZW model on $\Sigma$ with affine group $SU(N)_k$. The Hilbert space $\cH_{\Sigma}$ is finite dimensional and spanned by the finite number of conformal primaries in WZW model. In WZW the conformal primaries are finite in number and in one-to-one correspondence with integrable representations of affine Lie algebra $su(N)_k$. The first goal of this paper is to show that using the form of modular transform matrices $\mathcal{S}$ and $\mathcal{T}$ for given representations the partition function can be written as a unitary matrix model. These matrix models are similar to those studied by \cite{Aganagic:2002wv,Okuda}\footnote{The partition function for CS theories on different manifolds boils down to a novel class of matrix models\cite{Aganagic:2002wv}, with a Unitary Matrix Model(UMM) like measure and a non-periodic potential.} but with a difference. We observe that the eigenvalues of unitary matrices are proportional to hook numbers associated with an integrable representations. As a result the eigenvalues turn out to be discrete.

To check the consistency of our observation we consider the CS theory on $S^2\times S^1$ coupled with Gross-Witten-Wadia(GWW) potential\cite{gross-witten,wadia}\footnote{This partition function of \cs theory on $S^2\times S^1$ can also be written in terms of a sum over representations of $SU(N)$ \cite{Chattopadhyay,Chattopadhyay:2018wkp}. The $SU(N)$ representations are characterised by Young diagrams with maximum $N$ rows (no constraint on maximum number of columns). In our previous work \cite{Chattopadhyay:2018wkp} we observed that the discreteness in eigenvalue distribution imposes a constraint on dominant representations of $SU(N)$ : maximum number of columns must be less than $k$, which is nothing but the integrability condition. This observation motivated us to look at the relation between CS theories on different manifolds directly starting from its relation with the current algebra of the corresponding WZW theory. We postpone further discussion on this to conclusion section \ref{sec:conclu}.}. The resulting unitary matrix model turns out to be exactly same as the one derived earlier by \cite{shirazs2s1}\footnote{Using the technique developed in \cite{Blau:1993tv}. See also \cite{Blau:1993hj}} in the context of CS theory coupled with matter. The authors of \cite{shirazs2s1} also showed that the eigenvalues of unitary matrices are discrete and the discreteness in eigenvalues comes because of $U(1)$ fluxes through $S^2$. In our way of writing the partition function it seems that discreteness in eigenvalues emerges naturally. 

In a series of papers \cite{Naculich:1990,Naculich:1991,Naculich:1990hg,Nakanishi:1992,Naculich:2007nc} Naculich {\it et.al.} showed how the level-rank duality of WZW model flows to the level-rank duality of \cs theories. Under level rank duality a Wilson loop in one theory with some representation characterised by the Young diagram $Y$ maps to a Wilson loop with a representation $\tilde{Y}$ in the level-rank dual theory where $Y$ and $\tilde{Y}$ are related by transposition\footnote{Though this mapping is not in general one to one\cite{Naculich:2007nc}.}. Since the GWW model is self dual, writing down the partition function as a sum over integrable representations we explicitly check that different dominant representations at large $N$ are related to each other by transposition. In this paper we consider not only GWW potential, but also \cs theory of $S^2\times S^1$ coupled with fundamental matter. The partition function for such theory, in large $N$ limit, can be written as expectation value of an effective function of Wilson loops in pure \cs theory \cite{Aharony2012, shirazs2s1}. We express the partition function for these theories in terms of modular transformation matrices summed over integrable representations. This might help us understand the relation between \cs theory coupled with different fundamental matters using the transformation properties of modular transformation matrices under level-rank duality.

In this paper we also study pure level $k$, rank $N$ \cs theory on $S^3$. The partition function for this theory in canonical framing is trivial and given by $00$ component\footnote{$0$ representation means a Young diagram with no box.} of modular transformation matrix $\cS$. The  free energy at large $N$ matches with topological string theory on resolved conifold \cite{Gopakumar:1998ki} and exhibits no phase transition. The same partition function in \emph{Seifert framing} can be written as function of modular transform matrices summed over integrable representations. Using the properties of modular transform matrices one can show the equivalence between two framings up to a phase factor. However, the latter admits a matrix model representation of the theory. In this paper we show that the partition function can be written as a unitary matrix model where eigenvalues are discrete. The discreteness implies an upper cap in eigenvalue distribution in large $N$ limit. In our analysis we see that at large $N$ the partition function is dominated by an one-gap eigenvalue distribution that corresponds to an integrable representation and large $N$ free energy matches with the same in canonical framing. However, the large $N$ phase ceases to exist after a critical value of 't Hooft coupling $\lambda$. This implies that for 't Hooft coupling greater than the critical value the most dominant representation is not an integrable representation anymore. We do not have any satisfactory explanation for the existence of such critical value of 't Hooft coupling in the theory.

While studying large $N$ phases of pure \cs theory on $S^3$ we encounter another interesting phase of the model. We observe that for a range of 't Hooft coupling there exists a new phase of the system with similar free energy. The new phase corresponds to a two-gap distribution. Interestingly, the new phase (two-gap phase) also ceases to exist after the same critical value of 't Hooft coupling. However, we failed to find the new phase for lower values of 't Hooft coupling. This could be because of our limitation in numerical analysis. The bottom line is, for a finite range of 't Hooft coupling we observe that at large $N$ \cs theory on $S^3$ admits two topologically distinct eigenvalue distributions with same free energy. Understanding the physical meaning of these multigap phases in the topological string theory side \cite{Gopakumar:1998ki} is an interesting avenue to pursue.

The plan of this paper is as follows. In section \ref{sec:csonSM} we discuss how the partition function for \cs theory on Seifert manifold can be written as a function of modular transform matrices summed over integrable representations and their dependence on framings. We also show using the expressions for modular transform matrices one can write this partition function as a unitary matrix model. In section \ref{sec:css2s1} we consider \cs theory on $S^2\times S^1$ coupled with different fundamental matter. We write down the partition function in terms of modular transform matrices for any generic fundamental matter coupling. As a toy model we consider \cs theory coupled with GWW potential and find different dominant integrable representations for different phases of the theory at large $N$. From the dominant representations it is manifest that the theory is self-dual under level-rank duality. Section \ref{sec:csonS3} contains discussion of \cs theory on $S^3$. We show that at large $N$ there is a discrepancy in writing the partition function in Seifert framing. Namely, restriction on integrable representations seems to break down after a critical value of 't Hooft coupling. We summarise our main results in conclusion section \ref{sec:conclu} and discuss how the dominant representations found in the current paper are different than what we considered in our previous works \cite{Chattopadhyay,Chattopadhyay:2018wkp}.

\section{Chern-Simons partition function on Seifert manifold}
\label{sec:csonSM}

In this section we discuss how one can write the partition function (or correlation of Wilson loops) of a generic \cs theory on three dimensional compact manifold as a unitary matrix model.

We consider \cs theory of level $k$ (bare level) and gauge group $G(N)$ on a Seifert manifold. A Seifert manifold $\cM_{(g,p)}$ is a circle bundle over genus $g$ Riemann surface $\S_g$ with first Chern class $p$. Physical observables (Wilson loops) of \cs theory on such a manifold can be written in terms of observables in two dimensional WZW theory because of close connection between the two \cite{wittenjones}. A Seifert manifold for generic $p$ can be obtained from $\cM_{(g,0)}$ (which is a product of genus $g$ Riemann surface and a circle $\S_g\times S^1$) by surgery. Different choices of surgery give different framings of $\sm$. In a particular framing called Seifert framing, the expectation value of $n$ Wilson loops in different representations $\cR_1, \cR_2, \cdots, \cR_n$ of $G(N)$ can be written as \cite{Naculich:2007nc}
\ben\label{eq:WLonSM}
\cW_{\cR_1,\cdots,\cR_n}[\sm,G,k] = \sum_{\cR} (\cT_{\cR \cR})^{-p} \cS_{0\cR}^{2-n-2g} \prod_{i=1}^{n} \cS_{\cR\cR_i},
\een
where $\cT_{\cR\cR'}$ and $\cS_{\cR\cR'}$ are modular transform matrices that mix the affine characters associated with highest weight representations of affine Lie algebra $g(N)_k$ under translation and inversion of modular parameter $\tau$, respectively
\ben\label{eq:TSdef}
\chi_{\cR}(\tau+1) = \sum_{\cR' } \cT_{\cR \cR'} \chi_{\cR'}(\tau) \quad \text{and}\quad
\chi_{\cR}(-1/\tau) = \sum_{\cR'} \cS_{\cR \cR'} \chi_{\cR'}(\tau).
\een
All the sums in equation (\ref{eq:WLonSM}) and (\ref{eq:TSdef}) are over integrable representations of $g(N)_k$ and $\cR =0$ corresponds to identity representation. As mentioned earlier, we express all the representations in terms of Young diagram and hence an integrable representation means the Young diagram with maximum $k$ number of boxes in the first row (i.e. maximum $k$ columns).

For $G=\un$ the modular transform matrix $\cS_{\cR\cR'}$ of the $u(N)_k$ can be written in terms of modular transformation matrix of $su(N)$ and is given by\footnote{Affine Lie algebra of $\un$ WZW is the quotient of $su(N)_k \times u(1)_{N(k+N)}$ by $\mathbb{Z}_N$. Hence $u(N)$ representation can be written in terms of $su(N)$ representations and eigenvalues of $u(1)$ generator.}\cite{Naculich:1991}
\begin{eqnarray}
\begin{split}
\cS_{\cR\cR'}&=(-i)^{N(N-1)\over 2}(k+N)^{-N/2} e^{-{2\pi i Q Q' \over N(N+k)}}\det M(R,R').
\end{split}
\end{eqnarray}
We use the notation $\cR$ and $R$ for $u(N)$ and $su(N)$ representations respectively\footnote{One can follow  \cite{Dwivedi:2017rnj} for a generic discussion and derivation of modular transformation matrices of affine Lie algebra.}. $M(R,R')$ is a $N\times N$ matrix with elements,
\ben
M_{ij}(R,R') &= \exp\lB{2\pi i\over k+N}\phi_i(R)\phi_j(R')\rB
\een
where,
\ben\label{eq:phii}
\phi_i(R) = h_i-{s^R\over N}, \quad s^R=\sum_{i=1}^N h_i \quad \tand \quad h_i=n_i+N-i.
\end{eqnarray}
$n_i$'s are number of boxes in $i^{th}$ row of a given representation $R$ and $n_1\leq k$ with $R$ being an integrable representation. $Q$ is the eigenvalue of the $u(1)$ generator and is given by $Q= r(R)\ \text{mod} \ N$, $r(R)$ is the number of boxes in $R$.

After a little algebra, the $(0,\cR)$ component of modular transform matrix $\cS$ can be written in terms of hook numbers $h_i$'s,
\be
\cS_{0\cR} = (k+N)^{-N/2} 2^{N(N-1)/2} e^{-{2\pi i Q(\cR) Q(0) \over N(N+k)}} \prod_{i<j} \sin \lb{\pi(h_i-h_j)\over k+N} \rb.
\ee
$\cR=0$ means identity representation, i.e. $n_i=0, \ \forall\  i \in [1,N]$. As we will see, this expression plays an important role in our analysis.

The other modular transform matrix $\cT_{\cR\cR'}$ is given by
\ben
\begin{split}
	\cT_{\cR \cR'} =\expa{2\pi i(h_{R}-\frac{c}{24})} \delta_{\cR\cR'},\quad
	h_{R}=\frac{1}{2}\frac{C_{2}(\cR)}{k+N},\quad
	c =\frac{N(N k+1)}{k+N}
\end{split}
\een
and $\cC_2(\cR)$ is quadratic Casimir of $u(N)_k$
\begin{eqnarray}
\begin{split}
C_{2}(\cR)&= N\sum_{i=1}^N (l_i+s)+\sum_{i=1}^N (l_{i}+s)((l_{i}+s)-2i+1).
%\\
%&=\sum_{i}\left(h_{i}-\frac{(N-1-2s)}{2}\right)^{2}-\frac{N}{12}.
\end{split}
\end{eqnarray}
Here $s$ is any integer. Young diagram $\cR$ of $u(N)$ is obtained by prepending $s$ columns of $N$ boxes to the Young tableaux $R$ for the corresponding $SU(N)$. Thus the number of boxes in $\cR$ is given by $l_i+s$ and hence can be negative as well. However, we shall work in terms of number of boxes of the corresponding $su(N)$ representations $R$.

The partition function for \cs theory on $\cM_{(g,p)}$ can be written from equation (\ref{eq:WLonSM}) setting $n=0$,
\ben\label{eq:PFonSM}
\cZ_{CS}[\sm,U(N),k] = \sum_{\cR} (\cT_{\cR \cR})^{-p} \cS_{0\cR}^{2-2g},\quad \text{$\cR$ runs over integrable representations}. \ \ \ \ 
\een
We shall work with this partition function in this paper.

\subsection{Unitary matrix model and large $N$ limit}\label{sec:umm}

Starting with the above partition function (\ref{eq:PFonSM}), one can express the same in terms of hook number variables $\{h_i\}$,
\ben \label{eq:CSonSM2}
\begin{split}
\cZ_{CS}[\sm,G,k] &=  \lb {2^{(N-1)}\over k+N} \rb^{N(1-g)}   \sum_{\vec h} \ \prod_{i<j} \sin^{2-2g} \lb{\pi(h_i-h_j)\over k+N} \rb \\
& \qquad \qquad \qquad \qquad\qquad   e^{-\frac{i p \pi}{k+N} \sum_i (h_i-\Delta)^2 } e^{\frac{ \pi i p N^2}{12} -\frac{2(2-2g) i \pi  Q(R) Q(0)}{N (k+N)}} \\
\where \quad \Delta &= \frac12(N-1-2s)\ \ \tand \ \sum_{\vec h}\  = \sum_{\{h_i\}=0\atop h_1>h_2>\cdots>h_N}^{k+N} \ \text{is a restricted sum}.
\end{split}
\een
Here, $Q(R)= r(R)+s N$ and $Q(0)= s' N$, $r(R)$ being the number of boxes in $R$ representation. Since summation in equation (\ref{eq:CSonSM2}) is over the integrable representations, the hook number $h_i$ ranges between $0$ and $k+N$. To write the partition function in terms of a unitary matrix model we impose periodic boundary condition on $h_i \ : \ h_i \sim h_i+k+N$. We define angular variables $\{\theta_i\}$ as,
\be\label{eq:thetahrel}
\q_i = \lb{2\pi  \over k+N}\rb h_i,\quad \where \quad \q_N \geq 0 \quad \tand \quad \q_1\leq 2\pi.
\ee
Periodicity in $h_i$ implies $\theta_i \sim \theta_i+2\pi$. Here note that angular variables $\{\theta_i\}$'s are in monotonically decreasing order. In the large $N$ limit, the equation (\ref{eq:CSonSM2}) can be written in terms of redefined variables $\q_i$'s as
\ben \label{eq:CSonSM3}
\begin{split}
	\cZ_{CS}[\sm,G,k] &=   2^{N(N-1)}  \int \prod_i \frac{d\theta_i}{2\pi} \ \prod_{i<j} \sin^{2-2g} \lb{\theta_i -\theta_j\over 2} \rb e^{f(\{\theta_i\},k,N,p)}
\end{split}
\een
with some effective potential $f(\theta_i,k,N,p)$ depending on angular variable $\{\theta_i\}$ and other parameters. Thus, we see that the partition function can be written as a unitary matrix model with an effective potential $f(\{\theta_i\},k,N,p)$. The modular transform matrix $\cS_{0\cR}$ provides the correct measure factor for unitary matrix model.

From the redefinition (\ref{eq:thetahrel}) we see that that eigenvalues are discrete (since hook numbers can take only integer values). The discreteness of eigenvalues for \cs theory on $\stwoso$ was discussed in \cite{shirazs2s1} and the source of the discreteness was the $U(1)$ flux through $S^2$. Here we see that the discreteness is automatic for any three manifold (not just $\stwoso$) when we write \cs partition function as sum over integrable representations of $u(N)_k$ WZW model.

Discreteness in eigenvalues implies an upper cap in eigenvalue distribution function\footnote{We use a {\emph{negative}} sign in the definition of eigenvalue density because we arrange the eigenvalues in equation (\ref{eq:CSonSM2}) in monotonically decreasing order.} defined as,
\be\label{eq:evdensity}
\r(\q)=- \lim\limits_{\D x\ra 0} {\D x\over \D \q}, \quad \where \quad x=i/N \quad \tand \quad \q(x) =\q_i .
\ee
In the large $N$ limit
\be\label{eq:evbound}
\r(\q) \leq {1 \over 2\pi\l}, \quad \where \quad \l = {N\over k+N}.
\ee
In large $N$ limit, we can also define a Young diagram distribution function $u(h)$
\be\label{eq:uhdef}
u(h)= - \lim\limits_{\D x\ra 0} {\D x\over \D h}, \quad \where \quad x=i/N \quad \tand \quad h(x) =h_i/N.
\ee
Thus Young distribution and eigenvalue distributions are related by,
\ben\label{eq:urhorelation}
u(h)=2\pi \lambda \rho(2\pi \lambda h).
\een
From the relation $n_i=h_i-N+i$ we can independently check that the above $u(h)$ satisfies
\be
u(h)\leq 1
\ee
since, $n_i$'s are monotonically decreasing numbers. Therefore, this bound on $u(h)$ is consistent with the identification (\ref{eq:thetahrel}) and the bound on eigenvalue density (\ref{eq:evbound}). Thus we see that in the large $N$ limit, finding the most dominant Young diagram is equivalent to solving the saddle point equation in unitary matrix model. We should note that redefined variables $\theta_i$'s are periodic after we impose periodic boundary condition on hook numbers $h_i$.

\section{\cs theory on $\stwoso$ coupled with fundamental matter field}
\label{sec:css2s1}

In this section we write down the partition function of $U(N)$ Chern-Simons theory of level $k$ on $\stwoso$ coupled with different fundamental fields in terms of modular transform matrices of affine Lie algebra of boundary WZW model and study the properties of large $N$ representations that dominate the partition function. The actions for such matter couplings have been considered in many papers \cite{Jain:2013gza,Jain:2012qi,Jain:2013py,Giombi:2011kc,Minwalla:2011ma,Marino:2012az,Codesido:2014oua}.

It was shown in \cite{Aharony2012} that by integrating out the massive modes, thermal partition function of large $N$ \cs theory on $\stwoso$ can be written in terms of vacuum expectation value of effective action of holonomy $U$ along the thermal circle $S^1$,
\ben
\cZ_{CS}^{\text{matter}}\lB\stwoso, U(N),k\rB = \langle e^{-T^2 V_2 v(U)}\rangle_{N,k}
\een
where $T$ is the temperature (inverse of the size of the thermal circle), $V_2$ is the volume of $S^2$ and the effective potential $v(U)$ depends on the matter coupling.

The effective potential $v(U)$ can be computed case by case. In most of the cases, one can see that $v(U)$ can be written as a generic single plaquette model given by
\begin{equation}\label{plaquette}
\cZ(\beta)=\int \cD U\,\, \exp\left[N\sum_{n=1}^\infty{\beta_n(\beta)\over n}\left(\Tr U^n+\Tr U^{\dagger n}\right)\right]
\end{equation}
where $\beta_n(\beta)$'s are the parameters of the model. We now explicitly spell out how to write CS theory coupled with matter in the fundamental representation as an effective single plaquette model for four cases namely CS coupled to \emph{regular bosons, regular fermions, critical bosons} and \emph{critical fermions}. \\

\begin{itemize}
\item \textit{Chern-Simons coupled to regular bosons}

Consider Chern-Simons theory coupled to massless fundamental bosons or regular bosons with $\phi^6$ interaction \cite{giombi}. The action is given by
\begin{equation}
\nonumber
S=S_{CS}+S_{RB}
\end{equation}
where $S_{RB}$ is the matter action and is given by
\ben\label{eq:regularboson}
S_{RB} = \int d^3x \lB \lb D_\m \phi \rb^\dagger \lb D^\m \phi \rb  + {\lambda_6 \over 3! N^2} \lb \phi^\dagger \phi\rb^3 \rB .
\een
with $\lambda_{6}$ being the marginal coupling constant. The effective potential $v[\r]$ for this theory as a functional of eigenvalue density function $\rho(\theta)$ is given by \cite{shirazs2s1}
\begin{eqnarray}\label{eq:fundbose}
v[\r]= -{N\over 6\pi} \lb 1+{2\over \hat \l}\rb \s^3 +\frac{N}{2\pi} \int_{-\pi}^{\pi} d\theta \,\r({\theta}) \int_{\s}^{\infty} dy \,y \lb \ln(1-e^{-y +i\theta})+\ln(1-e^{-y-i\theta})\rb \nonumber \\
\end{eqnarray}
where
\be
\hat \l = \sqrt{{\l_6 \over 8\pi^2} +\l^2}
\ee
and $\lambda=N/(N+k)$ is the 't Hooft coupling.

Using the following expansion (since $|e^{-y+i\theta}|<1$ for $y>0$)
\be
\ln(1-e^{-y +i\theta}) + \ln(1-e^{-y -i\theta}) = -\sum_{n=1}^{\infty} \frac{e^{-n y}}{n} (e^{i n \theta}+e^{-i n \theta}),
\ee
and integrating over $y$ equation \ref{eq:fundbose} can be recast as,
\ben
v[U]= -\frac{N}{6\pi} \lb 1+{2\over \hat \l}\rb \s^3  -\frac{1}{2\pi}\sum_{n=1}^{\infty} {\b_n\over n} \lb \Tr U^n + \Tr U^{\dagger n}\rb
\een
where
\be
\b_n = \frac{e^{-n \s}(1+n\s)}{n^2}.
\ee

The value of $\s$ is obtained from minimization of $v[U]$ with respect to $\s$ in a given phase
\ben
\lb 1+{2\over \hat \l}\rb \s = \frac{1}{N}\sum_{n=1}^{\infty} \frac{e^{-n\s}}{n}\lb \r_n+\r_{-n}\rb,
\een
where $\rho_n= \Tr U^{n}$.

\item\textit{Chern-Simons coupled to regular fermions}

Consider now a theory of single massless fundamental fermion minimally coupled to a $U(N)$ level $k$ Chern-Simons theory. The action is given by
\begin{equation}
\nonumber
S=S_{CS}+S_{RF}
\end{equation}
where the matter action $S_{RF}$ is given by \cite{Giombi:2011kc}
\begin{equation}
S_{RF}=\int d^{3}x\hspace{3pt} \overline{\psi}\gamma^{\mu}D_{\mu}\psi.
\label{regularfermion}
\end{equation}
Following the same steps as in the case of fundamental bosons, the effective potential $v[\rho]$ given in \cite{shirazs2s1,Aharony2012} can be written as
\begin{equation}
v[U]=-\frac{N}{6\pi}\lb\frac{\tilde{c}^{3}}{\lambda}-\tilde{c}^{3}\rb-\frac{1}{2\pi}\sum_{n=1}^{\infty}\frac{\beta_{n}}{n}(TrU^{n}+TrU^{\dagger n})
\end{equation}
where
\begin{equation}
\beta_{n}=\frac{(-1)^{n+1} }{n^{2}}(1+n\tilde{c})e^{-n\tilde{c}}.
\end{equation}
The value of $\tilde{c}$ is determined by extremizing $v[\rho]$ with respect to $\tilde{c}$
\begin{equation}
\tilde{c}(\frac{1}{\lambda}-1)=-\frac{1}{N}\sum_{n=1}^{\infty}\frac{(-1)^{n}e^{-n\tilde{c}}}{n}(\rho_{n}+\rho_{-n}).
\end{equation}

\item\textit{Chern-Simons coupled to critical bosons}

At large $N$, critical bosonic theory can be defined as the Legendre transform of regular bosonic theory (equation (\ref{eq:regularboson}) with $\lambda_{6}=0$) with respect to the operator $\phi^{\dagger}\phi$. The action of regular boson theory is deformed by a mass squared parameter $A$ so that the action becomes
\begin{equation}
\nonumber
S=S_{Scalar}+S_{CS}+\delta S
\end{equation}
where
\begin{equation}
\delta S=\int d^{3}x\hspace{3pt} A\phi^{\dagger}\phi.
\end{equation}
The effective potential \cite{shirazs2s1} can be written as
\begin{equation}
v[U]=-\frac{N}{6\pi}\sigma^{3}-\frac{1}{2\pi}\sum_{n=1}^{\infty}\frac{\beta_{n}}{n}(TrU^{n}+TrU^{\dagger n})
\end{equation}
where
\begin{equation}
\beta_{n}=\frac{e^{-n\sigma}}{n^{2}}(1+n\sigma).
\end{equation}
Extremising the effective potential with respect to $\sigma$, we get
\begin{equation}
\sigma=\frac{1}{N}\sum_{n=1}^{\infty}\frac{e^{-n\sigma}}{n}(\rho_{n}+\rho_{-n}).
\end{equation}

\item\textit{Chern-Simons coupled to critical fermions}

Chern-Simons theory coupled to massless critical fermions in the fundamental representation can be defined as a deformation of regular fermion theory (Eq.(\ref{regularfermion})). The action is given by
\begin{equation}
S=S_{CS}+\int d^{3}x \hspace{3pt} \overline{\psi}\gamma^{\mu}D_{\mu}\psi+\int d^{3}x \hspace{3pt} (B\overline{\psi}\psi+\frac{N}{6}\lambda_{6}^{f}B^{3})
\end{equation}
where $B$ is a Lagrange multiplier field and $\lambda_{6}^{f}$ is a marginal coupling in the critical fermion theory.
The effective potential of the theory \cite{Jain:2013py} can be written as
\begin{equation}
v[U]=-\frac{N}{6\pi\lambda}\tilde{c}^{3}(1-\lambda+\hat{g}(\lambda,\lambda^{f}_{6}))-\frac{1}{2\pi}\sum_{n=1}^{\infty}\frac{\beta_{n}}{n}(TrU^{n}+TrU^{\dagger n})
\end{equation}
where
\begin{equation}
\beta_{n}=\frac{(-1)^{n+1}}{n^{2}}(1+n\tilde{c}) e^{-n\tilde{c}}.
\end{equation}
$\tilde{c}$ is obtained by extremizing the effective potential
\begin{equation}
\tilde{c}(\frac{1}{\lambda}-1+\frac{\hat{g}(\lambda,\lambda_{6}^{f})}{\lambda})=-\frac{1}{N}\sum_{n=1}^{\infty}\frac{(-1)^{n}e^{-n\tilde{c}}}{n}(\rho_{n}+\rho_{-n}).
\end{equation}
\end{itemize}

Thus, in all these examples we see that the partition functions take a generic form
\ben
\begin{split}
	\cZ_{cs}^{\text{matter}} & = \langle e^{v[U]}\rangle\\
	\where \quad v(U)& =N\sum_{n=1}^{\infty} \frac{\beta_n}{n} \lb \Tr U^n +\Tr U^{\dagger n}\rb +\beta_0.
\end{split}
\een
This way of writing the partition function is helpful in expressing the same in terms of modular transform matrices of $u(N)_k$ representations of dual WZW model.

\subsection{Connection with WZW model}

From the examples, discussed above, we see that coupling with fundamental matter fields renders an effective action of the form
\be
v(U)=N\sum_{n=1}^{\infty} \frac{\beta_n}{n} \lb \Tr U^n +\Tr U^{\dagger n}\rb +\beta_0,
\ee
where $\beta_n$ depends on the type of matter fields. Therefore the partition function of $U(N)$ \cs theory of level $k$ coupled with fundamental matter can be written in general as
\ben
\cZ_{CS}^{\text{matter}}\lB \stwoso, U(N),k\rB = \left<  e^{N\sum_{n=1}^{\infty} \frac{\beta_n}{n} \lb \Tr U^n +\Tr U^{\dagger n}\rb} \right>_{N,k}.
\een

Before we write down the partition function in terms of quantities of $u(N)_k$ WZW model, we make a justified approximation in $v(U)$. Since, $\beta_n$s are exponentially suppressed by a factor $e^{-\sigma n}$, we truncate the sum over $n$ and approximate that the sum runs from $1$ to $L$ where $L\leq k, N$. This enables us to write down the partition function in terms of modular transform matrices of $u(N)_k$ WZW model. However, in the limit $k\ra \infty$ and $N\ra \infty$ we can give up this approximation.

Expanding the exponential we can write,
\ben
\cZ_{CS}^{\text{matter}}\lB \stwoso, U(N),k\rB =  \sum_{\vec k, \vec l} g_{\vec k} \ g_{\vec l}\left< \Upsilon_{\vec k}(U) \Upsilon_{\vec l}(U^{\dagger}) \right>_{N,k}.
\een
where, $\vec \beta = \{\beta_1, \beta_2, \cdots,\beta_L\}$ is the set of $L$ parameters of the theory, $\vec k = \{k_1, k_2, \cdots, k_L\}$ and $\vec l = \{l_1, l_2, \cdots, l_L\}$ are $L$ dimensional vectors with $k_n, l_n \in [0,1,2,\cdots]$. Functions $g_{ \vec k}$ and $\Upsilon_{\vec k}(U)$ are given by,
\ben
\begin{split}
g_{\vec k} = \prod_{n=1}^{L}\frac{N^{k_n} \beta_n^{k_n}}{n^{k_n} k_n !}, \quad \Upsilon_{\vec k}(U) = \prod_{n=1}^{L} \lb\Tr U^n\rb^{k_n}.
\end{split}
\een
Using the group theory identity one can write
\ben
 \Tr U^m = \sum_{p=1}^{\text{min}(m,N)} (-1)^{p-1}\Tr_{R_p}U
\een
where the index $p$ in the representation $R_p$ defines different higher dimensional representations in terms of Dynkin indices as
\begin{eqnarray}
R_p=\left\lbrace\begin{matrix}
[m,0,\cdots,0] & \text{for }p=1\\
[\underbrace{m-p,0,\cdots,0,1}_p,0,\cdots,0] & \text{\quad for }p=2,\cdots,\text{min}(m,N-1)\\
[m-N,0,\cdots,0]& \text{for } p=N,m\geq N
\end{matrix}\right.
\end{eqnarray}
Since $n\leq k$, all the representations $R_p$ are integrable representations of $u(N)_k$. Clubbing up everything we get,
\ben
\begin{split}
\cZ_{CS}^{\text{matter}} &= \sum_{\vec k, \vec l} g_{\vec k} \ g_{\vec l}\prod_{n,m\atop =1}^L
\sum_{r_1,\cdots, r_{n}\atop s_1,\cdots,s_{m} } \frac{k_{n}!}{r_{1}!...r_{n}!}\frac{l_{m}!}{s_{1}!...s_{m}!}(-1)^{r_{2}+2r_{3}+...+(n-1)r_{n}+s_{2}+2s_{3}+...+(m-1)s_{m}}\\
& \qquad \qquad \langle (\Tr_{R_1} U)^{r_1} \cdots (\Tr_{R_{n}} U)^{r_{n}} (\Tr_{R_1} U^\dagger)^{s_1} \cdots (\Tr_{R_{m}} U^\dagger)^{s_{m}}\rangle_{N,k}.
\end{split}
\een
Using the definition (\ref{eq:WLonSM}) we finally write
\ben\label{eq:CSWZW}
\begin{split}
	\cZ_{CS}^{\text{matter}}\lB \stwoso, U(N),k\rB &= \sum_{\cR} \cS_{0\cR}^2 \exp\lB N\sum_{n=1}^{L} \frac{\beta_n}{n} \sum_{p=1}^{\text{min}(n,N)} (-1)^{p-1} \lb \frac{\cS_{\cR R_p} +\cS_{\cR\bar R_p}}{\cS_{0\cR}}\rb\rB. \ \ 
\end{split}
\een
In the limit $k,N \ra \infty$ we can lift the restriction over $n$ and all the $R_p$'s are integrable representations of $u(N)_k$. In that case, we can write the partition function of CS theory on $\stwoso$ coupled with fundamental matter in terms of quantities of related WZW model as,
\ben\label{eq:CSWZW2}
\begin{split}
	\cZ_{CS}^{\text{matter}}\lB \stwoso, U(N),k\rB &= \sum_{\cR} \cS_{0\cR}^2 \exp\lB N\sum_{n=1}^{\infty} \frac{\beta_n}{n} \sum_{p=1}^{\text{min}(n,N)}(-1)^{p-1} \lb \frac{\cS_{\cR R_p} +\cS_{\cR\bar R_p}}{\cS_{0\cR}}\rb\rB.\ \ \
\end{split}
\een
Transformation of modular transformation matrices under level-rank duality is well known \cite{Naculich:1990,Naculich:1990hg,Naculich:1991,Naculich:2007nc}. It would be interesting to see the level-rank duality between CS theory coupled with different fundamental matters using the dualities of modular transformation matrices.

\subsection{Gross-Witten-Wadia potential - A toy model}

Now we consider a toy example and discuss the dominant large $N$ representations corresponding to different phases of the theory. We also show how different large $N$ representations are related to each other by transposition as a consequence of level-rank duality.

We consider \cs theory on $\stwoso$ coupled with Gross-Witten-Wadia (GWW) potential. The partition function is given by 
\ben
\begin{split}\label{eq:gwwpf}
	\cZ_{CS}^{GWW} & = \int [\cD A] e^{S_{CS} + {N\b_1} (\Tr U+\Tr U^{\dagger})} 
	&= \dnorm{e^{{N\b_1} \lb \Tr U +\Tr U^{\dagger}\rb}}_{N,k}
\end{split}
\een
This particular example is self dual, since under level-rank duality the GWW potential transforms into itself \cite{shirazs2s1}. Using the relation (\ref{eq:CSWZW}) one can write the above partition function in terms of modular transform matrices
\be\label{eq:gwwpfwzw}
\cZ_{CS}^{GWW} = \sum_{\cR} \cS_{0\cR}^2 e^{{N\b_1} \lb {\cS_{\cR \cF} + \cS_{\cR\bar \cF}\over S_{0\cR}}\rb}.
\ee
Here $\cF$ and $\bar \cF$ stand for fundamental and anti fundamental representations. In large $N$ limit one can easily calculate that,
\ben
\begin{split}
	\frac{\cS_{\cR \cF}}{\cS_{\cR 0}} & =  e^{-2\pi i \phi}\sum_{i=1}^{N}e^{{2\pi i \over N+k} h_i } \quad \tand \quad  \frac{\cS_{\cR \bar \cF}}{\cS_{\cR 0}} & =  e^{2\pi i \bar \phi} \sum_{i=1}^{N}e^{-{2\pi i \over N+k} h_i}.
\end{split}
\een
where, $\phi = \frac{Q(R)(Q(\cF)-Q(0))+s^R}{N(k+N)}$ and $\bar\phi = \frac{Q(\cR)(Q(\cF)-Q(0))+s^R}{N(k+N)}$. By appropriately choosing the eigenvalues of $u(1)$ generators $Q$'s, we can set $\phi=0$ and $\bar\phi=0$. Hence the partition function is given by
\ben
\cZ_{CS}^{GWW} = 2^{N(N-1)} \lb {1\over k+N} \rb^N  \sum_{\vec h} \ \prod_{i<j} \sin^2 \lb{\pi(h_i-h_j)\over k+N} \rb e^{2 N\b_1 \sum_{i=1}^{N} \cosa{{2\pi \over N+k} h_i }}.
\een
In large $N$ limit we define,
\ben \label{eq:Nhdef}
h(x) = \frac{h_i}{N}, \quad \where \quad x=\frac i N, \quad x\in [0,1]
\een
with
\ben\label{eq:hxrange}
0\leq h(x)\leq \frac{1}{\lambda}.
\een
The partition function, in this limit, is given by
\ben
\cZ_{CS}^{GWW} = \cA \int [dh] e^{-N^2 \seff{h(x)}}
\een
where,
\ben
\begin{split}
	\seff{h(x)} &= - \Xint{-} dx dy \ln{|\sina{\pi \l (h(x)-h(y) )}|} - 2\b_1 \int dx \cosa{2\pi \l h(x)}.
\end{split}
\een
Redefining the variable $h(x)$, 
\ben
\q(x) = 2\pi \l h(x), \quad \with \quad 0\leq \q(x)\leq 2\pi
\een
the effective action in terms of $\q(x)$ can be written as,
\ben
\seff{\q(x)} = - \Xint{-} dx dy \ln{|\sina{\q(x)-\q(y)\over 2}|} - 2\b_1 \int dx \cos{ \q(x)}
\een
The saddle point equation in terms of eigenvalue density (equation \ref{eq:evdensity}) is given by,
\ben
\begin{split}
	\Xint{-} d\theta' \rho(\theta')\cot{\q-\q'\over 2} -2\b_1 \sin\q &=0.
\end{split}
\een
To find different large $N$ representations one has to solve this equation with the constraint $\rho(\theta) \leq \frac1{2\pi \lambda}$. This equation is exactly same as the eigenvalue equation discussed in \cite{shirazs2s1}. Therefore, in large $N$ limit, the dominant representations are completely determined by the corresponding dominant eigenvalue distributions studied in \cite{shirazs2s1}.

\subsubsection{Large $N$ representations}

In this section we study the dominant large $N$ integrable representations of $u(N)_k$ corresponding to \cs theory coupled with Gross-Witten-Wadia potential.

\subsubsection*{No-gap solution}

\begin{figure}[h]
	\centering
	\begin{subfigure}{0.4\textwidth}
		\centering
		\includegraphics[width=6.0cm,height=4cm]{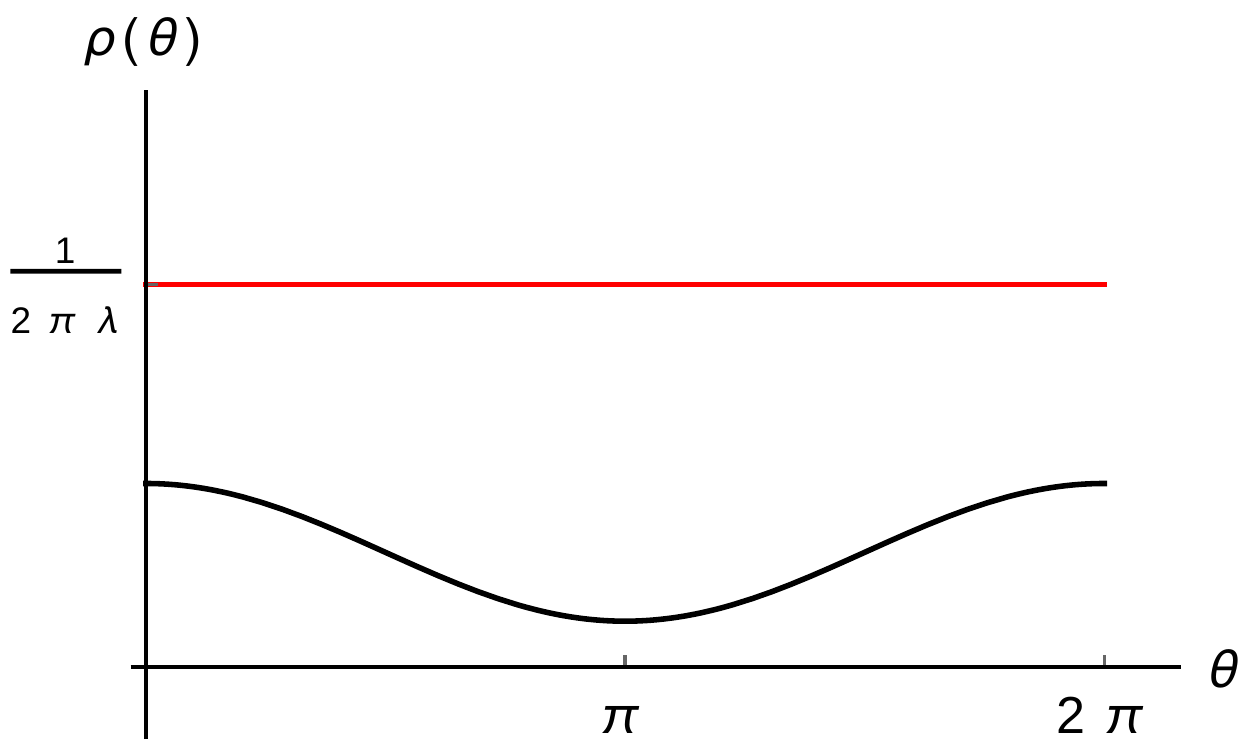}
		\caption{$\rho(\theta)$ vs. $\theta$ for no-gap phase (black curve). The red line denotes the upper-cap.}
	\end{subfigure}%
	\hspace{2cm}
	\begin{subfigure}{0.4\textwidth}
		\centering
		\includegraphics[width=5.0cm,height=3cm]{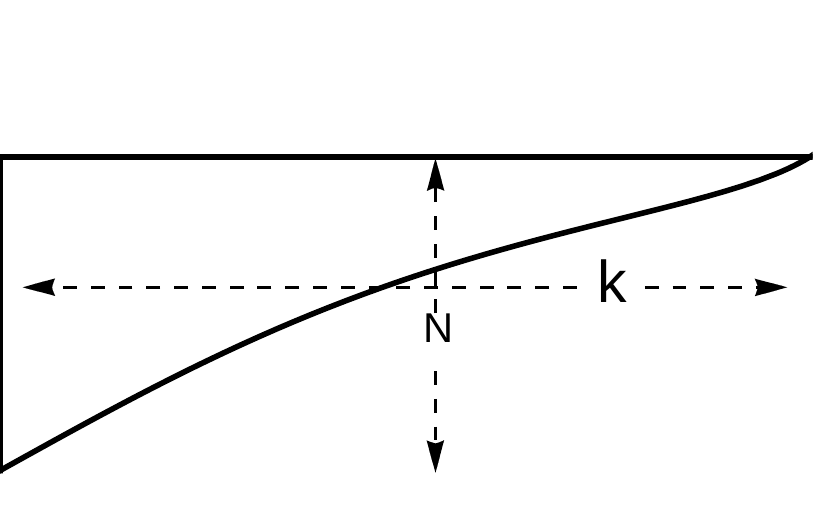}
		\caption{A typical Young diagram for no-gap phase. There are maximum $k$ columns. Young distribution function never saturates the upper bound 1.}
	\end{subfigure}
	\caption{Eigenvalue distribution and the corresponding dominant Young diagram for no-gap phase.}
	\label{fig:nogap}
\end{figure}

The no gap phase for capped GWW  model is identical with that of uncapped model. Eigenvalue distribution is given by
\begin{equation} \label{eq:evdistrinogap}
\rho(\gt)={1\over 2\pi}(1+2\gb_1\cos\gt).
\end{equation}
$\rho(\gt)$ is maximum (minimum) at $\gt=0, 2\pi \ (=\pi)$. Therefore from (\ref{eq:evbound}), we find that the no-gap phase is valid for
\ben\label{eq:nogapvalidity}
\begin{split}
	\beta_1<{1\over 2\gl}-{1\over 2}\quad &\text{for } \gl>{1\over 2}\\
	\beta_1<{1\over 2} \hspace{1.36cm} &\text{for }  \gl<{1\over 2}.
\end{split}
\een
In figure \ref{fig:nogap} we plot the eigenvalue density for $(\beta_1,\lambda)$ in the range mentioned in (\ref{eq:nogapvalidity}). 
%%%%%%%%%%%%%%%%%%%%%%%%%%%%%%%%%%%%%%%%%%%%%%%%%%
%%%%%%%%%%%%%%%%%%%%%%%%%%%%%%%%%%%%%%%%%%%%%%%%%%%%%

\subsubsection*{Lower-gap solution } 

Eigenvalue distribution for this phase is also same as the one-gap solution for uncapped GWW model \cite{gross-witten},
\begin{equation}\label{eq:evdistrionegap}
\begin{split}
\rho(\gt)&={2\gb_1 \over \pi}\sqrt{{1\over 2\beta_1}-\sin^2{\gt\over 2}}\,\,\,\left|\cos{\gt\over 2}\right|,\quad \text{for }\sin^2{\gt\over 2}<{1\over 2\gb_1}\\
\rho(\gt)&=0.\quad \text{for }\sin^2{\gt\over 2}>{1\over 2\gb_1}.
\end{split}
\end{equation}
The gap and distribution are distributed symmetrically around $\pi$. The maximum of this distribution is again at $\gt=0$. This phase only exists for $\gb_1\geq {1\over 2}$. Now we have further restriction due to upper limit of $\r(\q)$, which implies
\begin{eqnarray}\label{eq:onegapvalidity}
\gb_1\leq{1\over 8\gl^2}.
\end{eqnarray}  
Thus, lower-gap solution exists for 
\begin{equation}
\gb_1<{1\over 8\gl^2} \quad \text{and} \quad \gl\leq\half{1}.
\end{equation}
For $\gl>\half{1}$ this solution does not exist.

\begin{figure}[h]
	\centering
	\begin{subfigure}{0.4\textwidth}
		\centering
		\includegraphics[width=6.0cm,height=4cm]{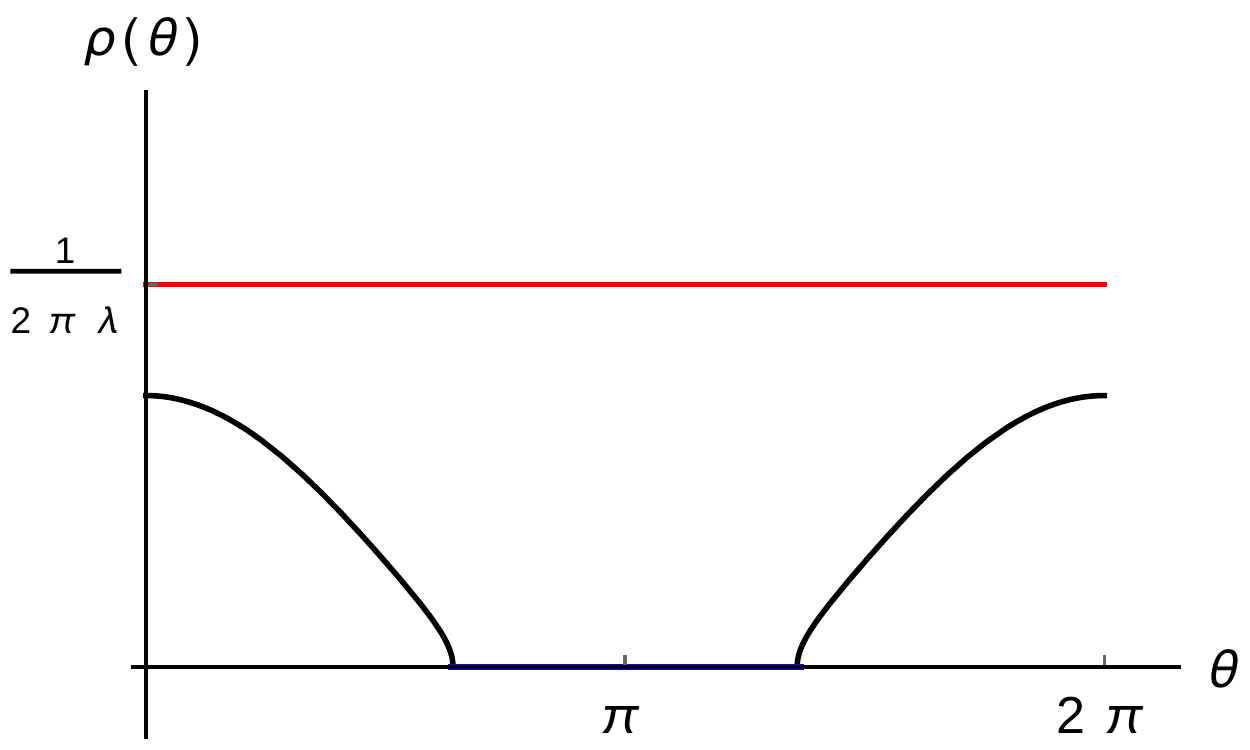}
		\caption{$\rho(\theta)$ vs. $\theta$ for lower-gap phase (black curve). The red line denotes the upper-cap. The blue line corresponds to gap in eigenvalue distribution.}
					\vspace{3.4cm}
	\end{subfigure}%
	\hspace{2cm}
	\begin{subfigure}{0.4\textwidth}
		\centering
		\includegraphics[width=5.0cm,height=3cm]{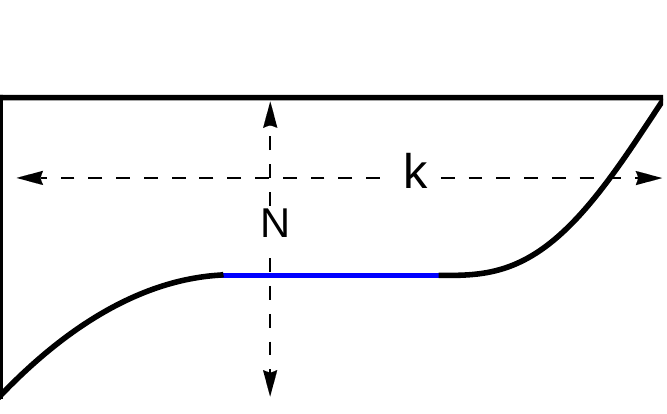}
	
		\caption{A typical Young diagram for lower-gap phase. There are maximum $k$ columns. $\rho(\theta)=0$ (equivalently $u(h)=0$) implies there is a horizontal jump in the Young diagram (blue line in the figure). Young distribution function never saturates the upper bound 1.}
	\end{subfigure}
	\caption{Eigenvalue distribution and the corresponding dominant Young diagram for lower-gap phase.}
	\label{fig:onegap}
\end{figure}

In figure (\ref{fig:onegap}) we plot the eigenvalue density for $(\beta_1,\lambda)$ in the range mentioned in (\ref{eq:nogapvalidity}). $\rho(\theta)=0$ (equivalently $u(h)=0$) implies that there is a horizontal jump in the Young diagram (blue line in the figure). 
%%%%%%%%%%%%%%%%%%%%%%%%%%%%%%%%%%%%%%%%%%%%%%%%%%
%%%%%%%%%%%%%%%%%%%%%%%%%%%%%%%%%%%%%%%%%%%%%%%%%%%%%

\subsubsection*{Upper-cap solution } 

This is the first new phase in capped GWW matrix model as well as any capped matrix models. In this phase though the eigenvalues are distributed like a no-gap solution, the distribution is saturated over some finite range. Following \cite{shirazs2s1} one can find eigenvalue density for upper cap solution as 
\begin{equation}\label{eq:evdistrionecap}
\begin{split}
\rho(\gt)&={1\over 2\pi\gl}-2\gb_1{|\sin {\gt\over 2}|\over \pi}\sqrt{{{{1\over \gl}-1}\over 2\gb_1}-\cos^2{\gt\over 2}}\quad \text{for }\cos^2{\gt\over 2}<{{{1\over \gl}-1}\over 2\gb_1}\\
\rho(\gt)&={1\over 2\pi\gl}\hspace{5.5cm} \text{for }\cos^2{\gt\over 2}>{{{1\over \gl}-1}\over 2\gb_1}.
\end{split}
\end{equation}
The minimum of this solution occurs at $\gt=\pi$ with the value 
$${1\over 2\pi}\left({1\over \gl}-2\sqrt{2\gb_1}\sqrt{{1\over \gl}-1}\right).$$
Now apart from being real, minimum value should also be greater than zero. Hence, this solution exists for 
\begin{equation}
{1\over 2\gl}-\half{1}<\gb_1<{1\over 8\gl(1-\gl)} \quad \text{for}  \quad \gl\geq{1\over 2}.
\end{equation}
The upper-cap solution does not exist for $\gl<\half{1}$.

\begin{figure}[h]
	\centering
	\begin{subfigure}{0.4\textwidth}
		\centering
		\includegraphics[width=6.0cm,height=4cm]{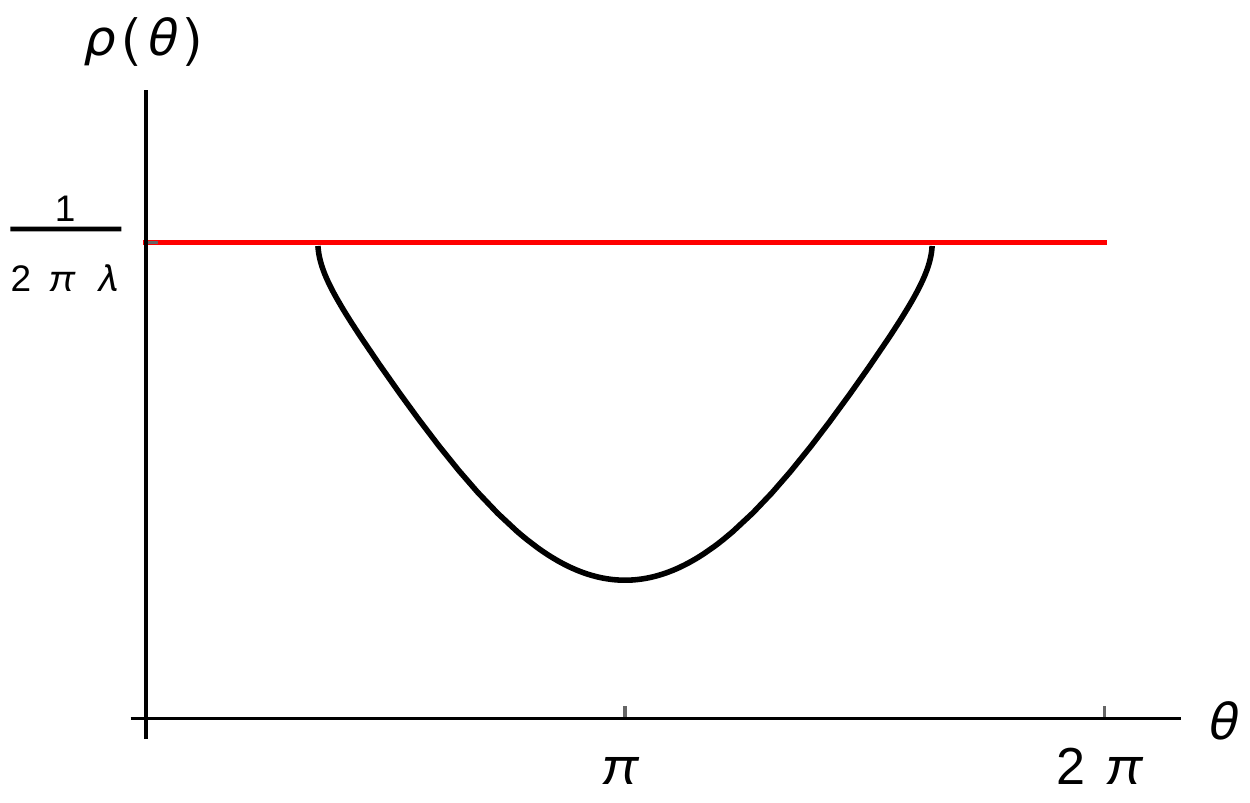}
		\caption{$\rho(\theta)$ vs. $\theta$ for upper-cap phase (black curve). The red line denotes the upper-cap.}
					\vspace{2.7cm}
	\end{subfigure}%
	\hspace{2cm}
	\begin{subfigure}{0.4\textwidth}
		\centering
		\includegraphics[width=5.0cm,height=3cm]{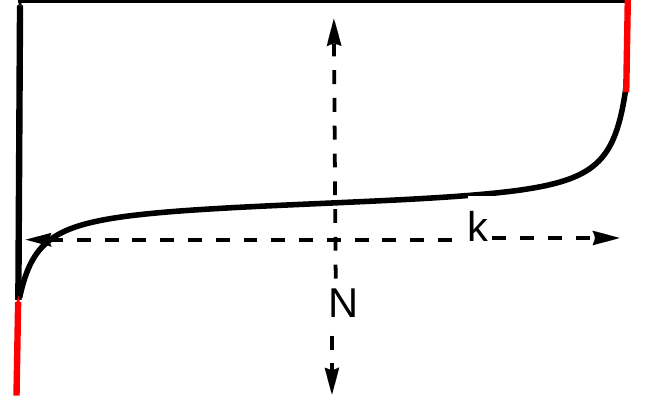}
		\caption{A typical Young diagram for upper-cap phase. There are maximum $k$ columns. Top red line corresponds to saturation of first few rows and the bottom red line implies last few rows are empty.}
	\end{subfigure}
	\caption{Eigenvalue distribution and the corresponding dominant Young diagram for upper-cap phase.}
	\label{fig:uppercap}
\end{figure}

In figure \ref{fig:uppercap} we see eigenvalue distribution as a function of $\theta$ and the corresponding Young distribution. The eigenvalue density touches the upper cap i.e. $1/2\pi\lambda$ in some range of $\theta$. This implies that the corresponding Young distribution function $u(h)$ touches 1. The Young distribution function touching unity implies that a finite fraction of rows have the same box numbers. Since $h=0$ corresponds to no box in the last row, this distribution implies a finite fraction of rows in the diagram are empty. Similarly, at the top a finite fraction rows have $k$ boxes in each (two red lines in the Young diagram).
%%%%%%%%%%%%%%%%%%%%%%%%%%%%%%%%%%%%%%%%%%%%%%%%%%
%%%%%%%%%%%%%%%%%%%%%%%%%%%%%%%%%%%%%%%%%%%%%%%%%%%%%

\subsubsection*{Upper-cap with lower-gap solution }

\begin{figure}[h]
	\centering
	\begin{subfigure}{0.4\textwidth}
		\centering
		\includegraphics[width=6.0cm,height=3.8cm]{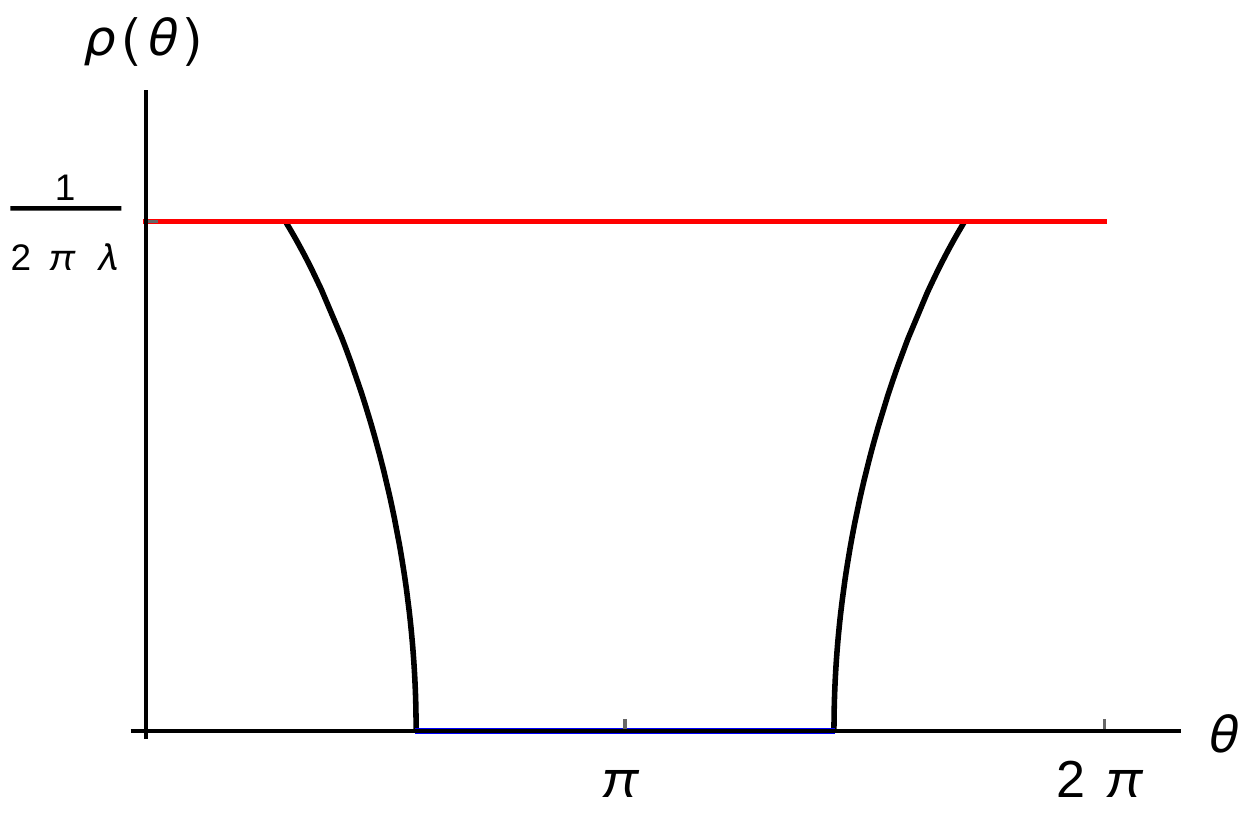}
		\caption{$\rho(\theta)$ vs. $\theta$ for cap-gap phase (black curve). The red line denotes the upper-cap.}
		\vspace{3.3cm}
	\end{subfigure}%
	\hspace{2cm}
	\begin{subfigure}{0.4\textwidth}
		\centering
		\includegraphics[width=5.0cm,height=3cm]{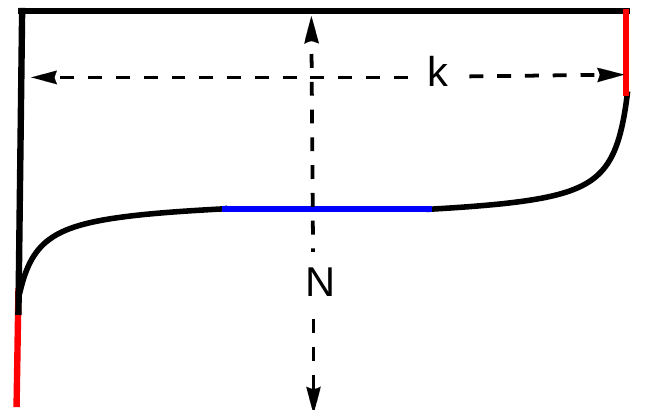}
		\caption{A typical Young diagram for cap-gap phase. Young diagram has a finite fraction of empty rows at the bottom and maximally saturated rows at the top. In between there are finite fraction of columns with same number of boxes.}
	\end{subfigure}
	\caption{Eigenvalue distribution and the corresponding dominant Young diagram for cap-gap phase.}
	\label{fig:capgap}
\end{figure}

The exact form of the solution for upper-cap with a lower gap is given in \cite{shirazs2s1}. Here we plot the corresponding eigenvalue density and the associated Young diagram in \ref{fig:capgap}. Here eigenvalue density touches both the upper cap $1/2\pi\lambda$ and the lower bound $0$. Hence, the corresponding Young diagram has a finite fraction of empty rows at the bottom and maximally saturated rows at the top. In between there are finite fraction of columns with same number of boxes. 
%%%%%%%%%%%%%%%%%%%%%%%%%%%%%%%%%%%%%%%%%%%%%%%%%%
%%%%%%%%%%%%%%%%%%%%%%%%%%%%%%%%%%%%%%%%%%%%%%%%%%%%%

\subsubsection{Level-Rank duality and transposition of diagrams}

The level-rank duality in terms of level $k$ and rank $N$ is given by $N\ra k$ and $k\ra N$. As a result, under level-rank duality, the 't Hooft coupling constant $\l$ transforms as
\be
\label{eq:paramdual1}
\gl^D={k\over k+N}=1-\gl, \quad \text{$\gl^D$ is the 't Hooft coupling in dual theory.}
\ee
Demanding that the partition function is invariant under level-rank duality we find that the second coupling constant $\bo$ also transforms under level-rank duality as
\be
\label{eq:paramdual2}
\gb_1^D={\gl\over 1-\gl}\gb_1, \quad \text{$\gb_1^D$ is the coupling in dual theory.}
\ee

It was shown in \cite{shirazs2s1} that under level-rank duality the eigenvalue densities for lower-gap and upper-cap phase are related to each other by,
\begin{eqnarray}\label{eq:rhodual}
\tilde\rho\left(\theta\right)={\gl\over 1-\gl}\left[{1\over 2\pi\gl}-\rho(\theta+\pi)\right].
\end{eqnarray}
From the relation between $\rho(\theta)$ and $u(h)$ (equation \ref{eq:urhorelation}) we see that under level-rank duality, Young distributions are related by
\ben
\tilde u(h) = \frac{\lambda}{1-\lambda}\lB 1- u\lb h+\frac1{2\lambda}\rb \rB.
\een
The above relation is a two step process. In the first step we shift the hook length by $\frac{1}{2\lambda}$ ($h\ra h +1/2\lambda$). Since the potential is periodic we can extend $h$ beyond $1/\lambda$ with the identification $h\sim h+ 1/\lambda$. 
%%%%%%%%%%%%%%%%%%%%%%%%%%%%%%%%%%%%%%%%%%%%%%%%%%
\begin{figure}[h]
	\centering
	\includegraphics[width=9.5cm,height=8.0cm]{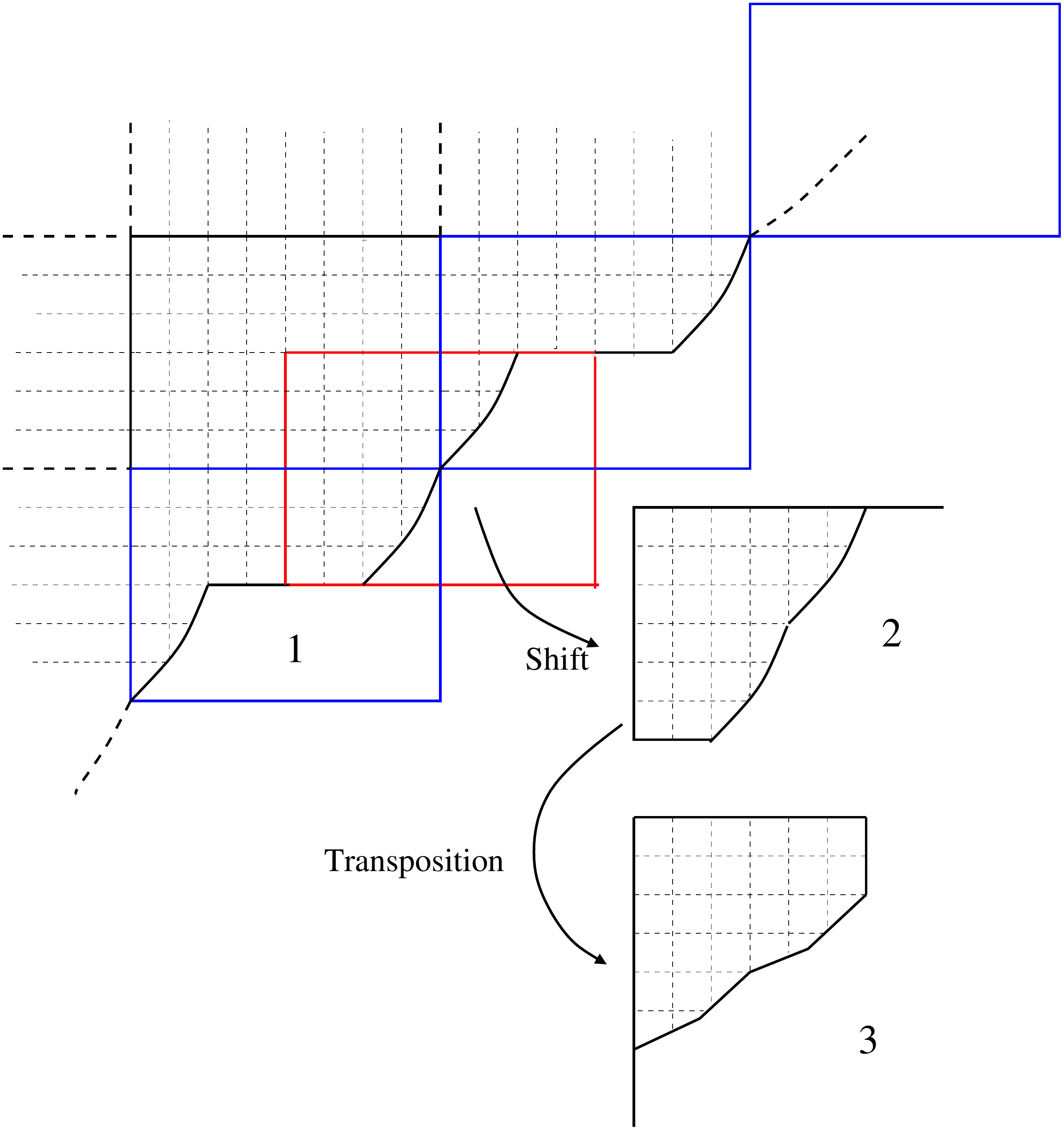}
	\caption{Duality in Young diagrams.}
	\label{fig:LRdualityYD}
\end{figure}
%%%%%%%%%%%%%%%%%%%%%%%%%%%%%%%%%%%%%%%%%%%%%%%%%%%%%
One can, therefore, periodically stack Young diagrams one after one (blue boxes) in figure \ref{fig:LRdualityYD}.
A shift of $1/2\lambda$ in $h$ means we go from blue box to the red box. The next step is $u(h)\ra 1-u(h)$. This implies transposition of the shifted diagram, i.e., row and columns are interchanged. In figure \ref{fig:LRdualityYD} we start with Young diagram for one-gap phase (diagram 1). After giving a shift in $h$ we obtain diagram $2$. Young diagram 3 is obtained from diagram 2 by transposition. Diagram 3 is the dominant one for upper-cap phase (figure \ref{fig:uppercap}). 

Since no-gap and cap-gap phases are dual to themselves, we see from the corresponding dominant Young diagrams that the above set of operations gives us back the same Young diagrams.

\section{\cs theory on $S^3$}
\label{sec:csonS3}

A special class of Seifert manifold ($g=0, \ p=1$) is three sphere : ${\cal M}_{0,1} = S^3$. \cs partition function on $S^3$ can be obtained from the generic expression (\ref{eq:WLonSM})
\ben\label{eq:cspfS3}
    \begin{split}
	\cZ_{\text{CS}}[S^3,U(N),k] = \sum_{\cR}  \cS_{0\cR}^{2} \cT^{-1}_{\cR \cR},\quad \text{in Seifert framing}.
	\end{split}
\een
Using the properties of modular transform matrices\footnote{$S^2=1$, $(ST)^3=S^2=1$.} one can show that the partition function in Seifert framing is same as that of in canonical framing up to a phase factor (see equation \ref{eq:framedepend})
\ben\label{eq:cspfS3CF}
\cZ_{\text{CS}}[S^3,U(N),k] =\cS_{00} \quad \text{in canonical framing}.
\een
Partition function in Seifert framing can be written as a matrix model. In the large $N$ limit we shall see that one integrable representation (say $\tilde \cR$) dominates the partition function and the value of the partition function for that representation is same as $\cS_{00}$
\be
\cZ_{\text{CS}}[S^3,U(N),k] = \lim\limits_{N\ra \infty} \cS_{0\tilde \cR}^2 \cT^{-1}_{\tilde \cR\tilde \cR} = \cS_{00}.
\ee

Using the expressions for modular transform matrices in terms of Young diagram data, the partition function for \cs theory on $S^3$ can be written as a matrix model\footnote{We appropriately choose $u(1)$ eigenvalues such that in the large $N$ limit the partition function matches with that in canonical framing (\ref{eq:cspfS3CF}).}
\ben\label{eq:CSpf0}
\begin{split}
	\cZ_{\text{CS}}[S^3,U(N),k] & =  \lb {2^{(N-1)}\over k+N}  \rb^N  e^{-{i\pi g_s(N^3-N)\over 6}}  \sum_{\vec h } \prod_{i<j}  \sin^2 \lB \pi g_s(h_i-h_j)\rB  e^{- {i \pi  g_s} \sum_i (h_i-\D)^2 }\ \ \ \ \\
	\where, \quad  g_s & =\frac1{k+N} \ \text{is coupling constant and } \ \D = \frac12(N-1-2s)
\end{split}
\een

There are two possible ways to deal with this partition function. Replacing $g_s\ra -i g_s$ one can write this partition function as a Hermitian matrix model. This was studied by \cite{Arsiwalla:2005jb,douglas-kazakov} in the context of two-dimensional Yang-Mills theory\footnote{Also look at \cite{Marino:2004eq} for a review.}. We briefly review the analysis in Appendix \ref{app:dualpf}. Here, we use the similar technique (section \ref{sec:umm}) to write the partition function as a unitary matrix model \cite{Okuda,Chattopadhyay}. 
We take $\D=\frac{k+N}{2}$ with appropriate choice of $s$ and define the angular variables $\theta_i$
\be
\q_i = {2\pi h_i \over k+N}-\pi,\quad \q_N \geq -\pi \quad \tand \quad \q_1\leq \pi.
\ee
The summand in equation (\ref{eq:CSpf0}) can be written as a measure of $SU(N)$ (or $\un$), as described in section \ref{sec:csonSM}, with the fact that eigenvalues are now discrete. Discreteness in eigenvalues implies an upper cap in eigenvalue distribution function given by equation (\ref{eq:evbound}).

The partition function in large $N$ limit after a wick rotation in $g_s$ ($g_s\ra i g_s$) is given by
\ben\label{eq:pfev}
\begin{split}
	\zcst =	e^{{\pi g_sN^3\over 6}} \int \prod_i \frac{d\q_i}{2\pi} 
	\prod_{i<j} 2 \sin^2 \lb \frac{\q_i-\q_j}{2}\rb e^{-\frac1{4\pi g_s} \sum_{i=1}^{N}\q_i^2}.
\end{split}
\een
The potential is not periodic in this case. However, to write a unitary matrix model for \cs theory we introduce periodicity in $h_i : h_i \sim h_i + k+N$. This implies that the harmonic oscillator potential $\theta^2$ is repeated beyond $-\pi \leq \theta \leq \pi$. Writing the partition function as a unitary matrix model has an advantage. In the next section (section \ref{sec:cstwogap}) we see that at large $N$, the model also admits a two-gap solution with same free energy. Finding this new phase is easy in unitary matrix model.

\begin{figure}[h]
	\centering
	\begin{subfigure}{0.4\textwidth}
		\centering
		\includegraphics[width=6.0cm,height=4cm]{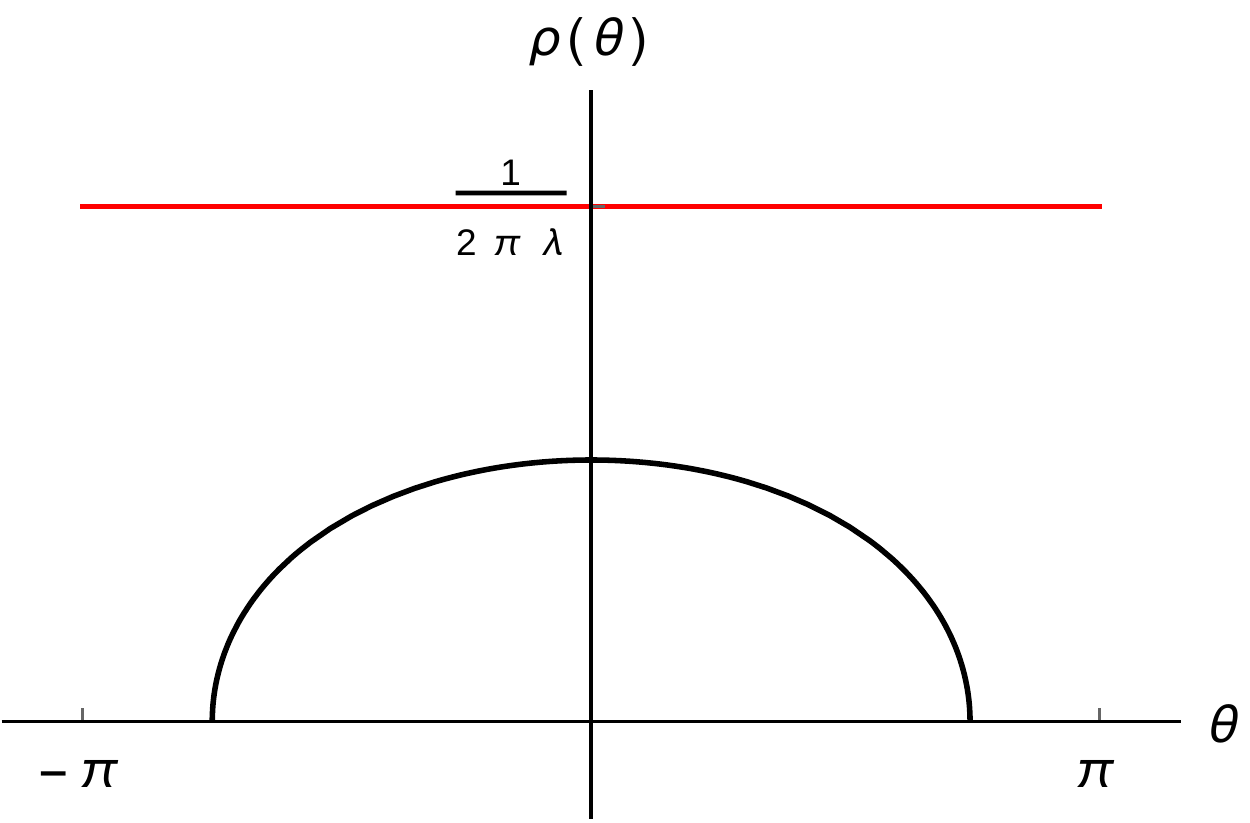}
		\caption{$\rho(\theta)$ vs. $\theta$ for one-gap phase (black curve). The red line denotes the upper-cap.}
		\vspace{.5cm}
	\end{subfigure}%
	\hspace{2cm}
	\begin{subfigure}{0.4\textwidth}
		\centering
		\includegraphics[width=5.0cm,height=3cm]{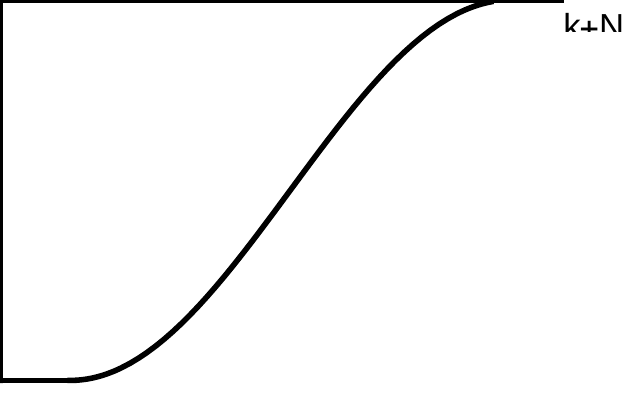}
		\caption{A typical Young diagram for one-gap phase.}
	\end{subfigure}
	\caption{Eigenvalue distribution and the corresponding dominant Young diagram for one-gap phase for CS on $S^3$.}
	\label{fig:CSonS3_nogap}
\end{figure}

In the large $N$ limit the eigenvalue distribution is governed by the saddle point equation
\ben
\begin{split}\label{eq:saddleeqnev}
	\Xint-d\q' \r(\q) \cot\lb \frac{\q-\q'}{2} \rb = \frac1{2\pi \l} \q.
\end{split}
\een
This unitary matrix model was studied in \cite{Chattopadhyay}. It was observed that the system has only one phase in the large $N$ limit and the eigenvalue distribution was given by,
\begin{eqnarray}\label{eq:csev1gap}
\r(\theta)&=& 
\frac{1}{2\pi^2 \l}\tanh^{-1}\left[ 
\sqrt{1- \frac{e^{-2\pi\l}}{\cos^{2}\frac{\theta}{2}}}\right].
\end{eqnarray}
Since $\r(\q)\geq 0$, this implies eigenvalues are distributed for $\theta \in [-2\cos^{-1}e^{-\pi\l}, 2\cos^{-1}e^{-\pi\l}]$. The eigenvalue distribution and corresponding Young diagram distribution are plotted in figure \ref{fig:CSonS3_nogap}.
%%%%%%%%%%%%%%%%%%%%%%%%%%%%%%%%%%%%%%%%%%%%%%%%%%
%%%%%%%%%%%%%%%%%%%%%%%%%%%%%%%%%%%%%%%%%%%%%%%%%%%%%
We call this phase a one-gap phase.

As discussed in \cite{Chattopadhyay} the theory does not admit any no-gap phase as the eigenvalue distribution for such phase becomes negative in some range of $\theta$. No phase transition in this theory was considered in \cite{Chattopadhyay} because eigenvalue density was unrestricted.

One can compute the free energy corresponding to this phase. The free energy is given by,
\ben\label{eq:css3free}
F_{CS}(\lambda)=\frac{2\pi^3 \lambda^3}{3} -\frac{\pi^3\lambda}{3}-Li_3(e^{-2\pi \lambda})+\zeta_3
\een
for all $0\leq\lambda\leq 1$. This is in agreement with \cite{Marino:2004eq, Gopakumar:1998ii,Periwal:1993yu,Gopakumar:1998ki}. This is, as expected, equal to the partition function (\ref{eq:cspfS3}) in canonical framing (\ref{eq:cspfS3CF}).

Interestingly, the eigenvalue distribution (\ref{eq:csev1gap}) is functional inverse of Young diagram distribution obtained in \cite{Marino:2004eq, Arsiwalla:2005jb} (see Appendix \ref{app:dualpf} for details). This identification is similar to that obtained in the context of a generic unitary matrix model considered in a series of papers \cite{duttagopakumar,duttadutta,Chattopadhyay}. However, in this case both the distributions correspond to large $N$ integrable representation of $u(N)_k$. This is similar to \cite{Grossmatytsin}.

From eigenvalue distribution (\ref{eq:csev1gap}), we see that the eigenvalue density saturates the upper-bound at $\lambda^*=\frac1\pi \ln \cosh\pi$. Saturation of eigenvalue density to its upper bound is equivalent to the dominant representation in the dual description, saturating the integrability condition. Therefore, one naturally expects that the system will undergo a phase transition from a gap phase to cap-gap phase after this value of $\lambda$ similar to \cite{shirazs2s1}. However, we explicitly check that such cap-gap phase does not exists for CS theory of $S^3$ which is manifest from the partition function written in canonical frame (no such phase transition exists in canonical frame). This essentially means that the representation $\tilde \cR$ which dominates the partition function in the large $N$ limit is not an integrable representation anymore. One needs to continue the sum over $\cR$ beyond integrable representations for $\lambda>\lambda^*$ \cite{Marino:2002fk, Rozansky:1994qe} using the symmetries of summand.

\subsection{Two-gap phase in \cs theory on $S^3$}
\label{sec:cstwogap}

Chern-Simons theory on $S^3$ does not allow any no gap solution\footnote{The saddle point equation does not allow a real semi-definite positive eigenvalue distribution for this case.} but admits multi-gap (more than one) solutions \cite{Morita2018,Chattopadhyay}. A two gap solution, in particular, was explicitly studied in \cite{Chattopadhyay}. The solution was given by,
\begin{eqnarray}\label{twogap}
\r(\theta) &=& \frac1{4\pi^2 \l} \tanh^{-1} \left[\frac{\sqrt{
		(2\cos\theta +\gamma)^2-4 e^{-4\pi\l} }}{2\cos\theta
	+\gamma} 
\right]\nonumber\\
&=& \frac1{4\pi^2 \l} \tanh^{-1}
\left[\frac{4\sqrt{\left(\cos^2{\theta/2}
		-\cos^2{\theta_1/2}\right)
		\left(\cos^2{\theta/2}-\cos^2{\theta_2/2}\right)}} 
{2\cos\theta +\gamma}\right],
\end{eqnarray}
where
\begin{equation}
\theta_1 = 2 \cos^{-1}
\left[\frac{\sqrt{2(1+e^{-2 \pi \l})-\gamma}}{2}\right],\quad\theta_2 = 2
\cos^{-1} 
\left[\frac{\sqrt{2(1-e^{-2\pi\l})-\gamma}}{2}\right].
\end{equation}
The parameter $\gamma$ can not be fixed from the analyticity or normalization conditions of resolvent. Hence it is an one parameter family of solutions at this point. However, the free energy for two-gap phase depends on the parameter $\gamma$ and hence, it is possible to cook up a new large $N$ solution or phase for a given value of $\lambda$ such that the free energy of that phase exactly matches with the free energy of one-gap phase. The additional parameter $\gamma$ depends on the 't Hooft coupling $\lambda$. Since the free energy matching condition is hard to track analytically we used numerical methods to find two-gap solutions with the same free energy as of the one-gap solution for a given value of $\lambda$. 

\begin{figure}[h]
%	\centering
	\begin{subfigure}{0.4\textwidth}
		\centering
	\includegraphics[width=7.5cm]{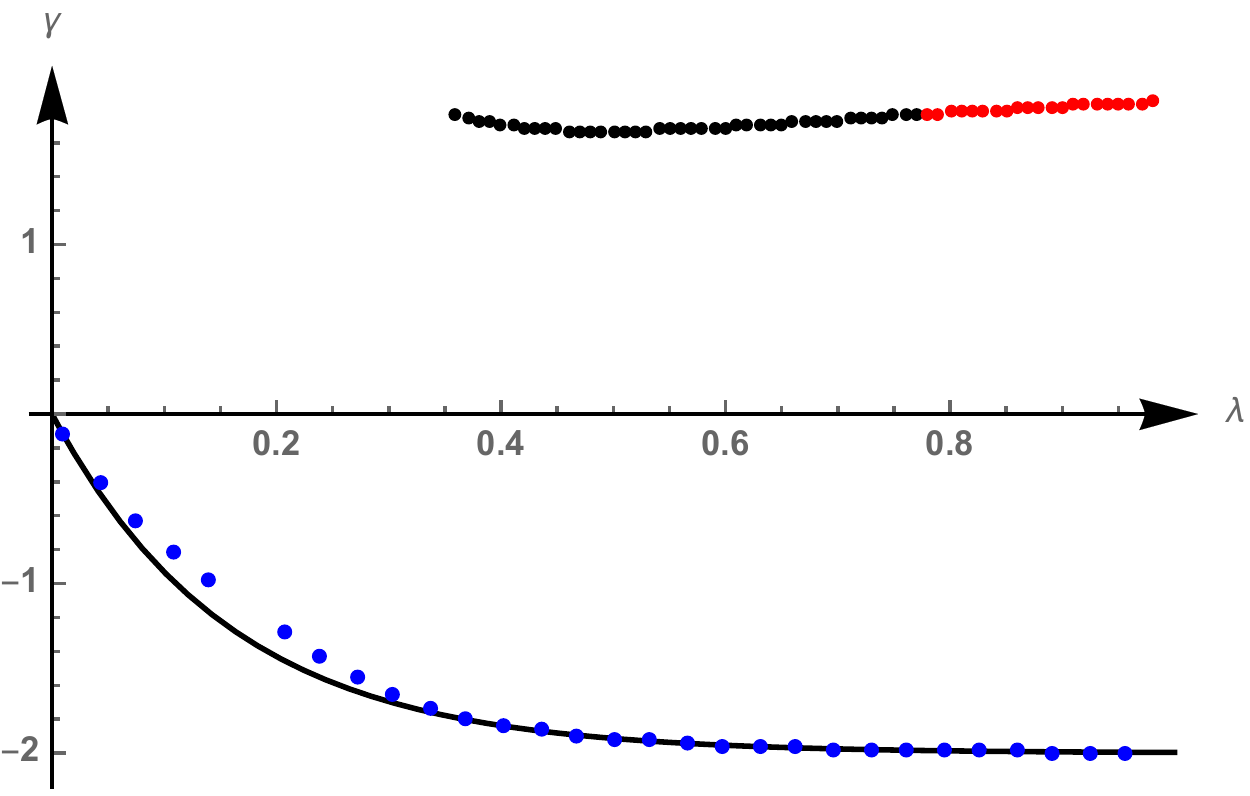}
	\caption{Numerical equi-energy plot for $\gamma$ vs $\lambda$.}
	\label{fig:gammall}
%	\vspace{.5cm}
\end{subfigure}%
\hspace{2cm}
\begin{subfigure}{0.4\textwidth}
	\centering
	\includegraphics[width=7.5cm]{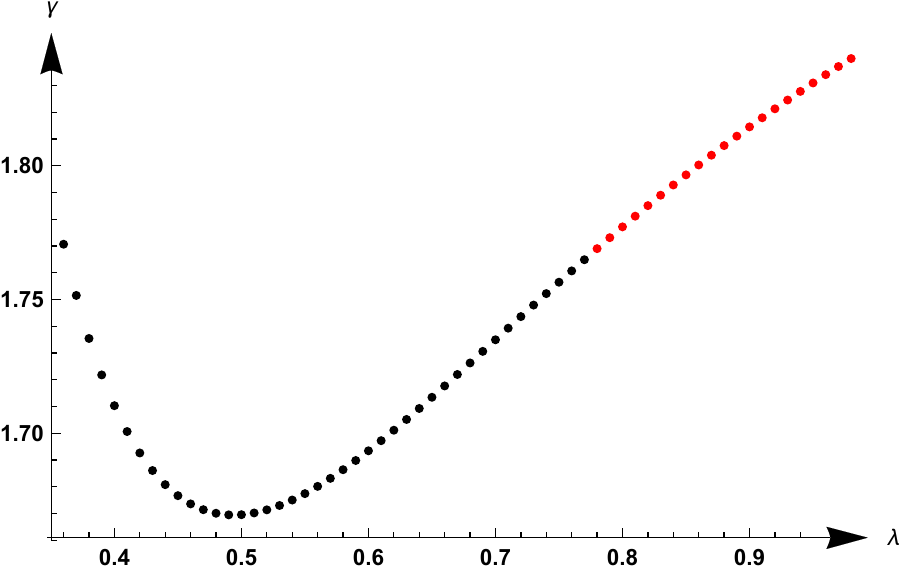}
	\caption{Zoomed in plot of $\gamma$ vs. $\lambda$ for two-gap phase (the upper graph of \ref{fig:gammall})}
\end{subfigure}
\caption{Equi-energy plot of $\gamma$ vs $\lambda$. Continuous lines denote analytic part and discrete points are the numerical findings. }
\end{figure}

The two-gap eigenvalue density (\ref{twogap}) is defined only when $\theta\in (-\pi,-\theta_1\}\cup\{-\theta_2,\theta_2\}\cup\{\theta_1,\pi\}$; therefore, it is quite clear if $\theta_1=\pi$ then the two gap solution should smoothly go to the one-gap solution and the free energies should match. This simple observation validates our numerical observations as well. One can observe from figure \ref{fig:gammall} (lower graph) that the analytic prediction (continuous line) indeed matches with the numerical prediction (discrete points in the figure). For all points on this line, the free energy of two-gap solution trivially matches with that of one-gap phase. 

\begin{figure}[h]
	\centering
	\begin{subfigure}{0.4\textwidth}
		\centering
		\includegraphics[width=6.0cm,height=4cm]{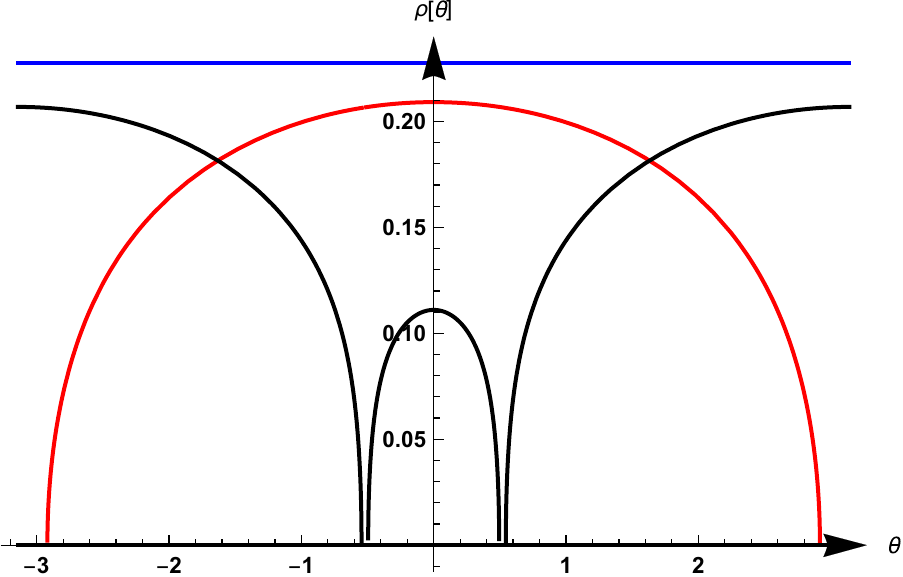}
		\caption{for $\lambda=0.7$ and $\gamma=1.73496$.}
		\vspace{.5cm}
	\end{subfigure}%
	\hspace{2cm}
	\begin{subfigure}{0.4\textwidth}
		\centering
		\includegraphics[width=6.0cm,height=4cm]{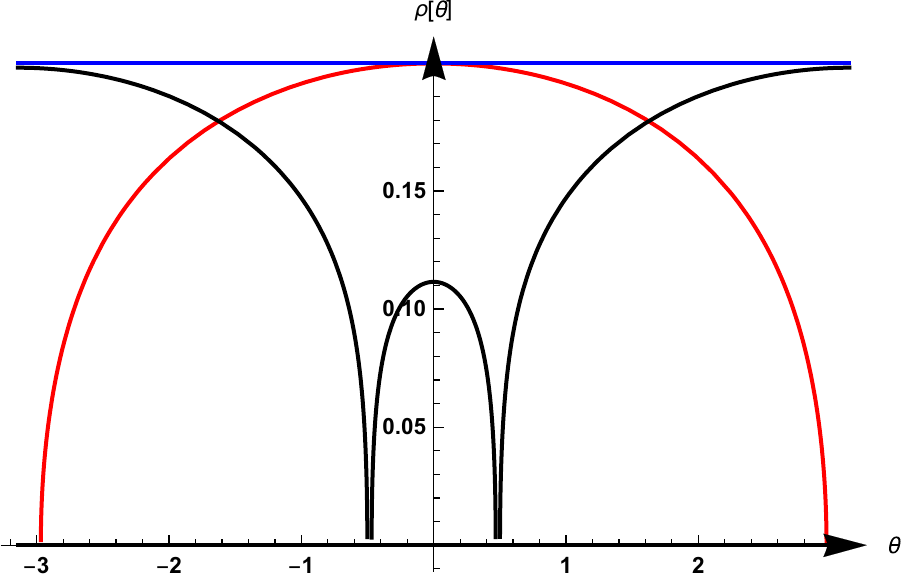}
		\caption{for $\lambda=\lambda^*$ and $\gamma=1.76900$.}
	\end{subfigure}\\

\begin{center}
		\begin{subfigure}{0.4\textwidth}
		\centering
		\includegraphics[width=6.0cm,height=4cm]{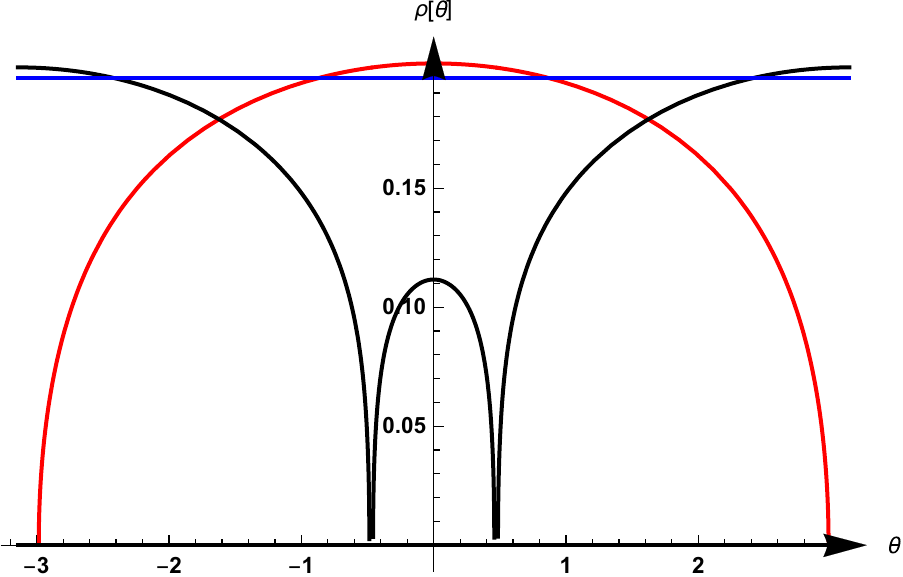}
		\caption{for $\lambda=0.81$ and $\gamma=1.78115$.}
	\end{subfigure}
\end{center}
 \caption{$\rho(\theta)$ vs. $\theta$ for one-gap phase (red curve), two-gap phase(black curve) and the upper saturation limit(blue line) for CS on $S^3$.}
	\label{fig:CSonS3full}
\end{figure}

In our numerical analysis we also find some non-trivial solutions (both $\theta_1$ and $\theta_2$ real, different and between 0 and $\pi$) of $\gamma$ as well for which the free energy matches with (\ref{eq:css3free}). We find that the non-trivial two-gap solution exists for $\lambda$ greater than a minimum value of $\lambda_0 = 0.3545 \pm .0045$. 
%%%%%%%%%%%%%%%%%%%%%%%%%%%%%%%%%%%%%%%%%%%%%%%
%%%%%%%%%%%%%%%%%%%%%%%%%%%%%%%%%%%%%%%%%%%%%%%
All the black points in figure \ref{fig:gammall} (upper graph) generate a two gap solution for which the free energy matches with the one-gap case. Surprisingly we observe that for $\lambda>\lambda^*$ this two-gap solution is also plagued with the same pathology {\it i.e.} eigenvalue density exceeding the value of upper-cap. For all the red points marked in the upper graph in figure \ref{fig:gammall}, one can generate a two-gap solution but for all of those points eigenvalue density goes beyond the capped value. In figure \ref{fig:CSonS3full} we plot eigenvalue distribution for one-gap and two-gap phase. For $\lambda>\lambda^*$, we see both one-gap and two gap solutions goes beyond the saturation limit. 

However, we are not very confident about the minimum value of $\lambda$ {\it i.e.} $\lambda_0$. Numerically we were not able to find any real acceptable values of $\theta_1$ and $\theta_2$ below $\lambda_0$ such that the free energy matches with one-gap phase. This could be because of our lack of expertise in numerical analysis. 
%%%%%%%%%%%%%%%%%%%%%%%%%%%%%%%%%%%%%%%%%%%%%%%%%%
%%%%%%%%%%%%%%%%%%%%%%%%%%%%%%%%%%%%%%%%%%%%%%%%%%%%%

 One of the primary goal of this numerical exercise is to establish the fact that in principle one can find a two-gap solutions\footnote{and also multi-gap solutions in the same spirit.} for CS theory on $S^3$ which has the same free energy as the one-gap phase. It should be emphasised that our objective in this section is to demonstrate the existence of multi-gapped phases in the matrix model side not to study any phase transition. As of now, these two different distributions mean two different integral representations for the observables of the theory. Although the underlying meaning of these multi-gap phases (condensation of $D2$ brane instantons ? \cite{Morita2018}) is not very clear yet, we attempt to show that such phases can be engineered with free energy same as that of one-gap phase. Understanding of physical meaning of these multi-gap phases (in topological string theory side) is an interesting avenue to pursue.

\section{Conclusion}
\label{sec:conclu}

In this paper we deal with a direct way of rewriting the partition functions of $SU(N)$ Chern-Simons theories on Seifert three manifolds as unitary matrix models and study the phase structure in the large $N$ limit. We start by considering the relation between expectation value of Wilson loops and modular transformation matrices of affine lie algebra pointed out by\cite{wittenjones}. Depending on a choice of framing one can write the partitions functions in a several possible equivalent ways. We chose a particular framing called the \emph{Seifert framing}. In this particular choice of framing the \cs partition functions can be written as function of modular transformation matrices of the corresponding affine lie algebra of the WZW model summed over highest weight representations. We show that by expressing the modular transformation matrices in terms of hook numbers of the corresponding integrable representations one can recover the unitary matrix models discussed in the literature of \cs theories on Seifert manifolds \cite{Marino:2002fk,Arsiwalla:2005jb,Okuda}. 

Our procedure naturally explains one crucial property that has been observed in \cs theories on $S^2\times S^1$ coupled with matter in the fundamental representation \cite{shirazs2s1}. It was observed in \cite{shirazs2s1} that, by carefully making correct gauge choices, one can write the full partition function as a UMM where eigenvalues of unitary matrices are discrete. In this paper we have seen that the discreteness in eigenvalues is universal for \cs theory on any 3-manifold that can be reached by doing surgery on $S^2\times S^1$. Here also we come across the notion of Young diagram density much in line with our previous works \cite{duttagopakumar,Chattopadhyay,Chattopadhyay:2018wkp,duttadutta}, but it should be noted that the Young diagram densities observed in this paper are different from that one discussed in our earlier works both in origin and interpretation. In our earlier works we expressed partition function of any $U(N)$ (or $SU(N)$) gauge theories (\cs theories in particular) as a sum of representations of $U(N)$, where there was no restriction on maximum number of columns. However, for \cs theory on \stso, in particular, we have seen that discreteness in eigenvalue distribution put a restriction on number of columns in $U(N)$ representations \cite{Chattopadhyay:2018wkp}. This restriction is equivalent to integrability condition on representation of affine group $U(N)_k$. In fact, it was shown in \cite{Chattopadhyay:2018wkp} that the eigenvalue distribution and Young diagram distribution for \cs theory on \stso coupled with GWW are functional inverses of each other and bear a meaning of free fermi description in the large $N$ limit. A clear hint was obtained from this work that partition function for \cs theory on \stso \ has a connection with integrable representations of affine Lie group $U(N)_k$. Motivated by that, in this paper we start with the result of \cite{wittenjones} and write partition function of \cs theory on \stso\ coupled with GWW (can be generalised to other fundamental matters) in terms of unitary matrix model. To our surprise, we see that at large $N$ Young diagram distributions are similar to eigenvalue distribution of \cite{shirazs2s1} (unlike functional inverse of the same as observed in \cite{Chattopadhyay:2018wkp}). An advantage of writing the partition function as a sum over integrable representation is that level-rank duality is manifest in terms of transposition of Young diagrams. We check this explicitly for GWW potential. Apart from answering the origin of our previous observation in \cite{Chattopadhyay:2018wkp}, this paper nicely explains the level-rank duality of \cs theories in a manifest way in certain sense. We have shown here that thermal partition functions of CS theory coupled with regular bosons or critical fermions on $S^2$ can be written as a effective single plaquette model and in principle one should be able to show the duality between these theories using the level-rank duality relation between modular transform matrices. 

As the manifold $S^3$ can be constructed out of $S^2\times S^1$ by means of surgery, it is no surprise that we find the same discreteness in eigenvalues in pure \cs theory on $S^3$ as well. Since CS theory on $S^3$ is purely topological one expects that there is no phase transition as shown earlier \cite{Marino:2002fk,Periwal:1993yu}. Therefore, the eigenvalue density should not saturate the upper bound. But surprisingly we found that at a particular value of the 't Hooft coupling defined as $\lambda^*$ in this text, the eigenvalue density saturates the maximum limit. However, no phase transition was observed at this value of 't Hooft coupling. We only understand that at large $N$ the dominant representation is not integrable any more after $\l>\l*$. Looking at the same problem through the canonical framing in place of Seifert framing it is clear that there is no such phase transition as shown by\cite{Periwal:1993yu}. It turns out that one can actually use the symmetry of the partition function to lift the constraint on the representation or in other words the constraint on the maximum number of columns in a representation and allow non-integrable representations to dominate the partition function as well \cite{Marino:2002fk,Rozansky:1994qe}. 

Using our earlier result \cite{Chattopadhyay}\footnote{Supported by numerical calculations of \cite{Morita2018}.} we  discuss about existence of a two-gap phase in \cs theory on $S^3$ and numerically tried to look for a solution\footnote{Though numerically challenging, but one can certainly look for such equi-free energy solutions for higher gap phases as well.} with the same free energy as that of the one-gap phase. But it turns out that even those solutions also saturates the upper value at $\lambda=\lambda^*$. We find that our two-gap solution breaks down for $\lambda$ less that a lower critical value $\lambda_0$. But this could be a pathology of our numerical analysis. The bottom line of our analysis is that at large $N$ there exists a two-gap phase for \cs theory on $S^3$ which has same free energy as one-gap phase for a finite range of 't Hooft coupling. This work also raises another interesting question about the existence of multi-gap phases for \cs theory on $S^3$. We are now trying to investigate the physical origin/meaning of this equi-free energy multigap phases in the context of its dual topological string theory\cite{Gopakumar:1998ki}.

\vspace{.6cm}
						
\noindent{\bf Acknowledgments:} We would like to thank Rajesh Gopakumar for many helpful discussion. We are grateful to S. Minwalla, D. Ghoshal, P. Dutta, R. Loganayagam, V. Singh, D. Mukherjee, S. Govindarajan for discussion. The work of SD is supported by the grant no.~EMR/2016/006294 from the Department of Science \&\ Technology, Government of India. SD and AC acknowledge the Simons Associateship of the Abdus Salam ICTP, Trieste, Italy. AC would like to thanks Robert de Mello Koch and all the organisers of third Mandelstam School and Workshop at Durban, where a part of the draft is written. AC would like to thank the hospitality of ICTP, Trieste where the final draft was prepared. We are indebted to people of India for their unconditional support towards researches in basic sciences. Finally we dedicate this paper to the brave souls of mother India who sacrificed their lives in \emph{Pulwama terror attack.}

%%%%%%%%%%%%%%%%%%%%%%%%%%%%%%%%%%%%%%%%%%%%%%%%%%%%%%%%%%%%%%%%%%%%%%%%%%%%%%%%
%%%%%%%%%%%%%%%%%%%%%%%%%%%%%%%%%%%%%%%%%%%%%%%%%%%%%%%%%%%%%%%%%%%%%%%%%%%%%%%%
%%%%%%%%%%%%%%%%%%%%%%%%%%%%%%%%%%%%%%%%%%%%%%%%%%%%%%%%%%%%%%%%%%%%%%%%%%%%%%%%
%%%%%%%%%%%%%%%%%%%%%%%%%%%%%%%%%%%%%%%%%%%%%%%%%%%%%%%%%%%%%%%%%%%%%%%%%%%%%%%%
%%%%%%%%%%%%%%%%%%%%%%%%%%%%%%%%%%%%%%%%%%%%%%%%%%%%%%%%%%%%%%%%%%%%%%%%%%%%%%%%
%%%%%%%%%%%%%%%%%%%%%%%%%%%%%%%%%%%%%%%%%%%%%%%%%%%%%%%%%%%%%%%%%%%%%%%%%%%%%%%%

\appendix

\section{A dual matrix model description}\label{app:dualpf}

\begin{figure}[h]
		\centering
		\begin{subfigure}{0.4\textwidth}
			\centering
		    \includegraphics[width=6.0cm,height=4cm]{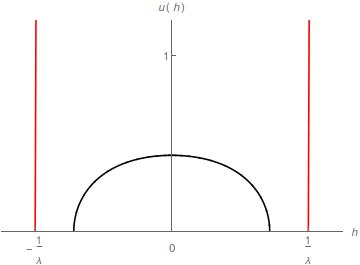}
		    \caption{$u(h)$ vs. $h$ for $\l<\l^*$}
		    \vspace{0.5cm}
		\end{subfigure}
	\hspace{2cm}
		\begin{subfigure}{0.4\textwidth}
		\centering
		\includegraphics[width=6.0cm,height=4cm]{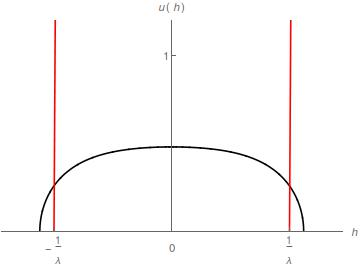}
		\caption{$u(h)$ vs. $h$ for $\l>\l^*$}
		\vspace{0.5cm}
	\end{subfigure}
		\caption{$u(h)$ vs. $h$ for different values of $\lambda$.}
		\label{fig:uhone_gap}
	\end{figure}

We start with equation (\ref{eq:CSpf0}).  Replacing $g_s\ra -i g_s$ one can write this partition function as a hermitian matrix model.
	\ben
	\begin{split}
		\zcst&=	\lb {2^{(N-1)}\over k+N} \rb^N e^{-{\pi g_s(N^3-N)\over 6}}
		\sum_{\vec{h}}  \prod_{i<j} \sinh^2 \lb \pi g_s(h_i-h_j) \rb e^{{-\pi g_s} \sum_i h_i^2}.
	\end{split}
	\een
	The partition function in large $N$ limit is given by,
	\ben\label{eq:CSpf1}
	\begin{split}
		\cZ_{cs}[S^3,U(N),k] &=  \cA(N,k) \int [\cD h] e^{-N^2 \seff{h}} 
	\end{split}
	\een
	where, 
	\ben\label{eq:Seffh}
	\begin{split}
		\seff{h} & = - \int dh u(h) \Xint{-}dh' u(h')\ln \bigg|\sinh \lB \pi \l (h-h'\rB\bigg| 
		+  \pi \l \int h^2 u(h) dh.
	\end{split}
	\een
	Here, $u(h)$ is Young diagram distribution function defined in equation (\ref{eq:uhdef}) with the constraint $u(h)\leq 1$. Although $h$ is a positive variable ranging between $0$ and $1/\lambda$, but effective action (\ref{eq:Seffh}) being an even function of $h$, one can extend the range of $h$ betwen $-1/\lambda$ to $1/\lambda$. In the large $N$ limit, the dominant contribution to partition function is determined by the saddle point equation
	\begin{eqnarray}\label{eq:saddleuh}
	\Xint-_{-1/\l}^{1/\l} dh'\,u(h') \coth\left(\pi \gl (h-h')\right)=h.
	\end{eqnarray}
	Solution of this equation is given by \cite{douglas-kazakov, Arsiwalla:2005jb}
	\ben\label{eq:uhsol}
	\begin{split}
		u(h) & =\frac{1}{\pi} \tan^{-1} \lB \sqrt{\frac{e^{2\pi\l}}{\cosh^2(\pi\l h)} -1 }\rB 
	\end{split}
	\een
	for  $-a\leq h\leq a$ and $0$ otherwise where $a= \frac{1}{\pi \l} \cosh^{-1}e^{\pi\l}$. It also turns out that for this solution $u(h)<1/2$ for $0\leq\l\leq1$. We call this \emph{one gap} solution as $u(h)=0$ in the complement region. This distribution represents a valid integrable representation for $a<1/\l$, which implies this phase is a valid phase of the theory for $0\leq \l < \l^* ,\ \where \ \l^*=\frac1{\pi}\ln \cosh\pi$. Young diagram distribution for $\l<\l^*$ and $\l>\l^*$ are plotted in figure \ref{fig:uhone_gap}.
	%%%%%%%%%%%%%%%%%%%%%%%%%%%%%%%%%%%%%%%%%%%%%%%%%%%%%%%
%%%%%%%%%%%%%%%%%%%%%%%%%%%%%%%%%%%%%%%%%%%%%%%%%%%%%%%%%
	As $\l$ increased beyond $\l^*$ the support of $h$ is greater than $1/\l$ hence this solution does not represent a valid integrable Young diagram.

	We take a pause and observe an interesting relation between eigenvalue distribution (\ref{eq:csev1gap}) and Young diagram distribution (\ref{eq:uhsol}) similar to \cite{Grossmatytsin}. They are functional inverse of each other
	\ben
	\begin{split} \label{eq:evYdrel}
		u(h)=\frac{\q}{2\pi}, \quad 2\r(\q) = \frac{h_+-h_-}{2\pi}
	\end{split}
	\een
	where $h_{\pm}$ are two roots of equation (\ref{eq:uhsol}). This identification is similar to that obtained in the context of a generic unitary matrix model considered in a series of papers \cite{duttagopakumar,duttadutta,Chattopadhyay} except for an extra 2 factor sitting in front of $\r(\q)$ in the second relation. Since $h_+-h_-$ has a maximum value $2/\l$, the eigenvalue distribution has an upper cap $\frac{1}{2\pi\l}$. All the eigenvalue distribution with $\r(\q)\leq \frac{1}{2\pi\l}$ correspond to integrable representations in the WZW side. It is easy to check that $\r(\q)$ saturates the upper cap at $\l=\l^*$. For $\l>\l^*$ we see that $h$ goes beyond $1/\lambda$ and hence the corresponding representation is not integrable representation. The above identification also tells that non-existence of \emph{no-gap} phase in eigenvalue side \cite{Chattopadhyay} is consistent with the fact that $u(h)$ can not have any capped phase \cite{Arsiwalla:2005jb}.

%%%%%%%%%%%%%%%%%%%%%%%%%%%%%%%%%%%%%%%%%%%%
%\bibliographystyle{hieeetr}
\bibliographystyle{hunsrt}
\bibliography{bibforcs}{}

\end{document}